\documentclass[aps,prd,twocolumn,groupedaddress]{revtex4}

  \usepackage{subfigure} 
\usepackage{graphicx}
\usepackage{epsf}

\newcommand{\sm}[1]{\mbox{{\scriptsize #1}}}

\newcommand{\bef}{\begin{figure}}
\newcommand{\eef}{\end{figure}}

\def\eps@scaling{0.95}

\def\showone#1{
  \centering
  \leavevmode
  \epsfxsize=\eps@scaling\linewidth
  \epsfbox{#1.eps}
}

\def\showtwover#1#2{
  \centering
  \leavevmode
  \epsfxsize=\eps@scaling\linewidth
  \epsfbox{#1.eps} \hfil
  \epsfxsize=\eps@scaling\linewidth
  \epsfbox{#2.eps}
}

\def\showthreeover#1#2#3{
  \centering
  \leavevmode
  \epsfxsize=\eps@scaling\linewidth
  \epsfbox{#1.eps} \hfil
  \epsfxsize=\eps@scaling\linewidth
  \epsfbox{#2.eps} \hfil
  \epsfxsize=\eps@scaling\linewidth
  \epsfbox{#3.eps}
}

\def\showfourover#1#2#3#4{
  \centering
  \leavevmode
  \epsfxsize=\eps@scaling\linewidth
  \epsfbox{#1.eps} \hfil
  \epsfxsize=\eps@scaling\linewidth
  \epsfbox{#2.eps} \hfil
  \epsfxsize=\eps@scaling\linewidth
  \epsfbox{#3.eps} \hfil
  \epsfxsize=\eps@scaling\linewidth
  \epsfbox{#4.eps}
}

\def\epstwo@scaling{0.46}

\def\showtwo#1#2{
  \centering
  \leavevmode
  \epsfxsize=\epstwo@scaling\linewidth
  \epsfbox{#1.eps} 
  \epsfxsize=\epstwo@scaling\linewidth
  \epsfbox{#2.eps}
}

\def\epsthree@scaling{0.28}
\def\showthree#1#2#3{
  \centering
  \leavevmode
  \epsfysize=\epsthree@scaling\textwidth 
  \epsfbox{#1.eps} 
  \epsfysize=\epsthree@scaling\textwidth 
  \epsfbox{#2.eps}
  \epsfysize=\epsthree@scaling\textwidth 
  \epsfbox{#3.eps}
}

\def\epstwo@scaling{0.44}

\def\showfour#1#2#3#4{
  \centering
  \leavevmode
  \epsfxsize=\epstwo@scaling\linewidth
  \epsfbox{#1.eps} \hfil
  \epsfxsize=\epstwo@scaling\linewidth
  \epsfbox{#2.eps} \hfil
  \epsfxsize=\epstwo@scaling\linewidth
  \epsfbox{#3.eps} \hfil
  \epsfxsize=\epstwo@scaling\linewidth
  \epsfbox{#4.eps}
}

\def\showsix#1#2#3#4#5#6{
  \centering
  \leavevmode
  \epsfxsize=\epstwo@scaling\linewidth
  \epsfbox{#1.eps} \hfil
  \epsfxsize=\epstwo@scaling\linewidth
  \epsfbox{#2.eps} \hfil
  \epsfxsize=\epstwo@scaling\linewidth
  \epsfbox{#3.eps} \hfil
  \epsfxsize=\epstwo@scaling\linewidth
  \epsfbox{#4.eps} \hfil
  \epsfxsize=\epstwo@scaling\linewidth
  \epsfbox{#5.eps} \hfil
  \epsfxsize=\epstwo@scaling\linewidth
  \epsfbox{#6.eps}
}

\newcommand{\befone}{
  \begin{figure*}
  \centering
  \begin{minipage}{\textwidth}
  }
\newcommand{\eefone}{\end{minipage}\end{figure*}}

\newcommand{\HI}{$\mathrm{H}$\ }
\newcommand{\HII}{$\mathrm{H}^+$\ }

\newcommand{\HzI}{$\mathrm{H}_2$\ }

\newcommand{\HeI}{$\mathrm{He}$\ }

\newcommand{\fHI}{\mathrm{H}}
\newcommand{\fHII}{\mathrm{H}^+}

\newcommand{\fHeI}{\mathrm{He}}

\newcommand{\fe}{\mathrm{e}}

\def\jnl@style#1{{\rmfamily#1}}%
\def\jref@jnl#1{{\jnl@style#1}}%

\newcommand\aj{\jref@jnl{AJ}}%
\newcommand\araa{\jref@jnl{ARA\&A}}%
\newcommand\apjl{\jref@jnl{ApJ}}%
\newcommand\apjs{\jref@jnl{ApJS}}%
\newcommand\apss{\jref@jnl{Ap\&SS}}%
\newcommand\aap{\jref@jnl{A\&A}}%
\newcommand\aapr{\jref@jnl{A\&A~Rev.}}%
\newcommand\aaps{\jref@jnl{A\&AS}}%
\newcommand\azh{\jref@jnl{AZh}}%
\newcommand\baas{\jref@jnl{BAAS}}%
\newcommand\jrasc{\jref@jnl{JRASC}}%
\newcommand\memras{\jref@jnl{MmRAS}}%
\newcommand\mnras{\jref@jnl{MNRAS}}%
\newcommand\pasp{\jref@jnl{PASP}}%
\newcommand\pasj{\jref@jnl{PASJ}}%
\newcommand\qjras{\jref@jnl{QJRAS}}%
\newcommand\skytel{\jref@jnl{S\&T}}%
\newcommand\solphys{\jref@jnl{Sol.~Phys.}}%
\newcommand\sovast{\jref@jnl{Soviet~Ast.}}%
\newcommand\ssr{\jref@jnl{Space~Sci.~Rev.}}%
\newcommand\zap{\jref@jnl{ZAp}}%
\newcommand\iaucirc{\jref@jnl{IAU~Circ.}}%
\newcommand\aplett{\jref@jnl{Astrophys.~Lett.}}%
\newcommand\apspr{\jref@jnl{Astrophys.~Space~Phys.~Res.}}%
\newcommand\bain{\jref@jnl{Bull.~Astron.~Inst.~Netherlands}}%
\newcommand\fcp{\jref@jnl{Fund.~Cosmic~Phys.}}%
\newcommand\gca{\jref@jnl{Geochim.~Cosmochim.~Acta}}%
\newcommand\grl{\jref@jnl{Geophys.~Res.~Lett.}}%
\newcommand\jgr{\jref@jnl{J.~Geophys.~Res.}}%
\newcommand\jqsrt{\jref@jnl{J.~Quant.~Spec.~Radiat.~Transf.}}%
\newcommand\memsai{\jref@jnl{Mem.~Soc.~Astron.~Italiana}}%
\newcommand\nphysa{\jref@jnl{Nucl.~Phys.~A}}%
\newcommand\physrep{\jref@jnl{Phys.~Rep.}}%
\newcommand\physscr{\jref@jnl{Phys.~Scr}}%
\newcommand\planss{\jref@jnl{Planet.~Space~Sci.}}%
\newcommand\procspie{\jref@jnl{Proc.~SPIE}}%

\begin{document}

\title{Dark stars: Implications and constraints from cosmic reionization and extragalactic background radiation}


\author{Dominik R. G. Schleicher, Robi Banerjee, Ralf S. Klessen}
\email{dschleic@ita.uni-heidelberg.de}
\affiliation{Institute of Theoretical Astrophysics / ZAH,\\ Albert-Ueberle-Str. 2,\\ D-69120 Heidelberg,\\ Germany}

\begin{abstract}
Dark stars powered by dark matter annihilation have been proposed as the first luminous sources in the universe. These stars are believed to form {in the central dark matter cusp} of low-mass minihalos. {Recent calculations indicate stellar masses up to $\sim1000\,M_\odot$ and/or have very long lifetimes. The UV photons from these objects could therefore contribute significantly to cosmic reionization.} Here we show that such dark star models would require a somewhat artificial reionization history, based on a double-reionization phase and a late star-burst near redshift $z\sim6$, in order to fulfill the WMAP constraint on the optical depth as well as the Gunn-Peterson constraint at $z\sim6$. This suggests that, if dark stars were common in the early universe, then models are preferred which predict a number of UV photons similar to conventional Pop.~III stars. This excludes $800\ M_\odot$ dark stars that enter a main-sequence phase and other models that lead to a strong increase in the number of UV photons.\\
We also derive constraints {for massive as well as light dark matter candidates} from the observed X-ray, gamma-ray and neutrino background, considering dark matter profiles which have been steepened during the formation of dark stars. This increases the clumping factor at high redshift and gives rise to a higher dark matter annihilation rate in the early universe. {{We furthermore estimate the potential contribution from the annihilation products in the remnants of dark stars, which may provide a promising path to constrain such models further, but which is currently still uncertain.}}
\end{abstract}


\maketitle

\section{Introduction}
Growing astrophysical evidence suggests that dark matter in the universe is self-annihilating. X-ray observations from the center of our Galaxy find bright $511$ keV emission which cannot be attributed to single sources \citep{Jean06, Weidenspointner06}, but can be well-described assuming dark matter annihilation \citep{BoehmHooper04}. {Further observations indicate also an excess of GeV photons \citep{deBoer05}, of microwave photons \citep{Hooper07}, and of positrons \citep{Cirelli08}. A common feature of these observations is that the emission seems isotropic and not correlated to the Galactic disk. However, there is usually some discrepancy between the model predictions and the amount of observed radiation, which may be due to uncertainties in the dark matter distribution, astrophysical processes and uncertainties in the model for dark matter annihilation \citep{deBoer08}.      }

{It is well-known that weakly-interacting massive dark matter particles may provide a natural explanation of the observed dark matter abundance \citep{Drees93, Kolb90}.} Calculations by \citet{Ahn05b} indicated that the extragalactic gamma-ray background cannot be explained from astrophysical sources alone, but that also a contribution from dark matter annihilation is needed at energies between $1$-$20$ GeV. {It is currently unclear whether this is in fact the case or if a sufficient amount of non-thermal electrons in active galactic nuclei (AGN) is available to explain this background radiation \citep{Inoue08}. Future observations with the FERMI satellite \footnote{http://www.nasa.gov/mission\_pages/GLAST/science/index.html} will shed more light on such questions and may even distinguish between such scenarios due to specific signatures in the anisotropic distribution of this radiation \citep{Ando07}. }

{The first stars have been suggested to have high masses of the order $\sim100\ M_\odot$, thus providing powerful ionizing sources in the early universe \citep{Abel02, Bromm04}. The effect of dark matter annihilation on the first stars has been explored recently in different studies. \citet{Spolyar08} showed that an equilibrium between cooling and energy deposition from dark matter annihilation can always be found during the collapse of the proto-stellar cloud. This has been explored further by \citet{Iocco08} and \citet{FreeseSpolyar08}, who considered the effect of scattering between baryons and dark matter particles, increasing the dark matter abundance in the star. \citet{IoccoBressan08} considered dark star masses in the range $5\leq M_*\leq600\ M_\odot$ and calculated the evolution of the pre-main-sequence phase, finding that the dark star phase where the {energy input} from dark matter annihilation dominates may last up to $10^4$~yr. \citet{FreeseBodenheimer08} examined the formation process of the star in more detail, considering polytropic equilibria and {additional mass accretion} until the total Jeans mass of $\sim800\ M_\odot$ is reached. They find that this process lasts for $\sim5\times10^5$~yr. They suggest that dark stars are even more massive than {what is} typically assumed for the first stars, and may be the progenitors for the first supermassive black holes at high redshift. {\citet{IoccoBressan08}, \citet{Taoso08} and \citet{Yoon08} have calculated the stellar evolution for the case in which the dark matter density inside the star is enhanced by the capture of addition WIMPs via off-scattering from stellar baryons. \citet{IoccoBressan08} followed the stellar evolution until the end of \HeI burning, \citet{Yoon08} until the end of oxygen burning and \citet{Taoso08} until the end of \HI burning. \citet{Yoon08} also took the effects of rotation into account.} The calculations found a potentially very long lifetime of dark stars and correspondingly a strong increase in the number of UV photons that may contribute to reionization. Dark stars in the Galactic center have been discussed by \citet{Scott08a, Scott08b}.
}

{Such models for the stellar population in the early universe imply} that the first luminous sources produce much more ionizing photons, and reionization starts earlier than for a population of conventional Pop.~III stars. In fact, we recently demonstrated that reionization based on massive Pop.~III can well reproduce the observed reionization optical depth \citep{SchleicherBanerjee08a}. Increasing the number of ionizing photons per stellar baryon may thus reionize the universe too early and produce a too large reionization optical depth. This can only be avoided by introducing a transition to a stellar population which produces less ionizing photons, such that the universe can recombine after the first reionization phase. We therefore consider a double-reionization scenario in order to re-obtain the required optical depth. We discuss such models in \S \ref{reionization} and demonstrate that some models of dark stars require considerable fine-tuning in reionization models in order to be compatible with the reionization optical depth from the WMAP \footnote{http://lambda.gsfc.nasa.gov/} 5-year data \citep{Nolta08, Komatsu08} and to complete reionization at redshift $z\sim6$ \citep{Becker01}. In \S \ref{21cm}, we show how such scenarios can be tested via $21$ cm measurements.

A further consequence of the formation of dark stars is the steepening of the density profiles in minihalos \citep{FreeseBodenheimer08, Iocco08}, thus increasing the dark matter clumping factor with respect to standard NFW models. In \S \ref{xray}, we estimate the increase in the clumping factor during the formation of dark stars and compare the calculation with our expectation for conventional NFW profiles {and heavy dark matter candidates. In \S \ref{light}, we perform similar calculations for the light dark matter scenario}. Further discussion and outlook is provided in \S \ref{outlook}.

\section{The models}\label{models}
As discussed in the introduction, various models have been suggested for dark stars. The main difference between these models comes from considering or neglecting scattering between dark matter particles and baryons. In addition, it is not fully clear how important a phase of dark matter capture via off-scattering from baryons actually is, depending on further assumptions on the dark matter reservoir. In the following, we will thus distinguish between main-sequence dominated models and capture-dominated models. 

\subsection{Main-sequence dominated models}
{After an initial phase of equilibrium between cooling and heating from dark matter annihilation \citep{Spolyar08, Iocco08, FreeseSpolyar08}, the dark star will contract further while the dark matter annihilates away and the heating rate thus decreases. This duration of this adiabatic contraction (AC) phase is currently controversial: While \citet{IoccoBressan08} find it to be in the range of $(2-20)\times10^3$~yr, \citet{FreeseBodenheimer08} require about $10^6$~yr. However, with a surface temperature of $\sim6000$~K, the stars are rather cold in this phase, and thus will not contribute significantly to reionization. The uncertainty in the duration of the AC phase is therefore not crucial in this context.}

{If the elastic scattering cross section as well as the dark matter density around the star are sufficiently large, the star will enter a phase which is dominated by the capture of further dark matter particles. Such a scenario will be discussed in more detail in the next subsection. Here, we assume that the elastic scattering cross section is either too small, or that the dark matter reservoir near the star is not sufficient to maintain the capture phase for long. Then, the star will enter the main-sequence phase (MS), in which the luminosity is generated by nuclear burning. Stars with $\sim1000\ M_\odot$ are very bright in this phase, and emit $\sim4\times10^4$ hydrogen-ionizing photons per stellar baryon during their lifetime \citep{Bromm01, Schaerer02}. We will refer to stars of such type, which have only a short or even no phase driven by dark matter capture, as MS-dominated models.}

{For the case of MS-dominated models, we will focus essentially on the very massive stars suggested by \citet{FreeseBodenheimer08}. For stars in the typical Pop.~III mass range, it has been shown elsewhere \citep[e. g.][]{SchleicherBanerjee08a} that they are consistent with reionization constraints. A star with $\sim800\ M_\odot$ forming in a dark matter halo of $\sim10^6\ M_\odot$ corresponds to a star formation efficiency of $1\%$, which we adopt for this case.
}

\subsection{Capture-dominated models}
{For a non-zero spin-dependent scattering cross section between baryons and dark matter particles, stars can capture additional WIMPs which may increase the dark matter density inside the star. For a cross section of the order $5\times10^{-39}$~cm$^2$ and an environmental dark matter density of $\sim10^{10}$~GeV~cm$^{-3}$, this contribution becomes significant and alters the stellar evolution during the main sequence phase.}  We will refer to such a scenario as a capture-dominated (CD) model. These phases have been studied in detail by {\citet{IoccoBressan08},} \citet{Taoso08} and \citet{Yoon08}. They found that the number of ionizing photons produced by such stars may be considerably increased with respect to high-mass stars without dark matter annihilation effects, which is mostly due to a longer lifetime. In particular for dark matter densities of $(1-5)\times10^{10}$~GeV~cm$^{-3}$ , the number of produced ionizing photons may be increased by up to two orders magnitude, while it decreases rapidly for larger dark matter densities, and the number of ionizing photons per baryon even drops below the value for Pop.~II stars at.threshold densities of $1\times10^{12}$~GeV~cm$^{-3}$. As \citet{Yoon08} found only a weak dependence on stellar rotation, we will not explicitly distinguish between models with and without rotation in the follow.

{ For the calculation of reionization, we will focus on some representative models of \citet{Yoon08} in the following. However, we point out that there are still significant uncertainties in these models, in particular the dark matter parameters and the lifetimes of the stars. The latter should be seen as upper limits, as they assume that a sufficient reservoir of dark matter is available in the stellar neighborhood to allow for ongoing dark matter capture. This may however be disrupted by dynamical processes. An apparent disagreement of dark stars in the early universe with our reionization model may thus indicate that the stellar lifetimes are indeed smaller due to such processes. }

\section{Reionization constraints}\label{reionization}
In this section, we briefly review our reionization model and discuss reionization histories for main-sequence and capture-dominated models. {These calculations implicitly assume annihilation cross sections of the order $10^{-26}\ \mathrm{cm}^2$ and dark matter particle masses of the order {$100$} GeV, the values which are typically adopted in dark star models}. In such models, dark matter annihilation does not contribute to cosmic reionization \citep{SchleicherBanerjee08a}. The chemistry in the pre-ionization era is thus unchanged and well-described by previous works \citep{Puy93, Galli98, Stancil98, Schleicher08}, such that the initial conditions for star formation are unchanged. Considering higher annihilation cross sections essentially yields an additional contribution to the reionization optical depth, which would sharpen the constraints given below.

\subsection{General approach}
Our calculation of reionization is based on the framework developed by \citet{SchleicherBanerjee08a}, which we have implemented in the RECFAST code \footnote{http://www.astro.ubc.ca/people/scott/recfast.html} \citep{Seager99,Seager00}. We will review here only those ingredients which are most relevant for this work. During reionization, the IGM consists of a two-phase medium, i.~e. a hot ionized phase and a rather cold and overally neutral phase. The relative size of these phases is determined from the volume-filling factor $Q_{\fHII}$ of the \HII regions \citep{ShapiroGiroux87,Haiman98,Barkana01,Loeb01,Choudhury05,Schneider06} as a function of redshift, given by
\begin{equation}
 \frac{dQ_{\fHII}}{dz}=\frac{Q_{\fHII}C(z)n_{\fe,\fHII}\alpha_A}{H(z)(1+z)}+\frac{dn_{\sm{ph}}/dz}{n_{\fHI}},
\end{equation}
where $C(z)=27.466\ \mathrm{exp}(-0.114z+0.001328z^2)$ is the clumping factor \citep{Mellema06}, $n_{\fe,\fHII}$ the number density of ionized hydrogen, $\alpha_A$ the case A recombination coefficient \citep{Osterbrock89}, $H(z)$ the Hubble function, $n_{\fHI}$ the mean neutral hydrogen density in regions unaffected by UV feedback and $dn_{ph}/dz$ the UV photon production rate. Our model consists of ordinary differential equations (ODEs) for the evolution of temperature $T$ and ionized fraction $x_i$ in the overall neutral medium. For the application considered here, the dominant contribution to the effective ionized fraction $x_{eff}=Q_{\fHII}+(1-Q_{\fHII})x_i$ and the effective temperature $T_{eff}=10^4\ \mathrm{K}\ Q_{\fHII}+T(1-Q_{\fHII})$ comes indeed from the UV feedback of the stellar population, i. e. from the hot ionized phase. According to \citet{Gnedin98} and \citet{Gnedin00}, we introduce the filtering mass scale as
\begin{equation}
M_{F}^{2/3}=\frac{3}{a}\int_0^a da' M_J^{2/3}(a')\left[1-\left(\frac{a'}{a} \right)^{1/2} \right],\label{filter}
\end{equation}
where $a=(1+z)^{-1}$ is the scale factor and $M_J$ the thermal Jeans mass, given as
\begin{equation}
M_J=2M_\odot \left(\frac{c_s}{0.2\ \mathrm{km/s}} \right)^3\left(\frac{n}{10^3\ \mathrm{cm}^{-3}} \right)^{-1/2}.
\end{equation}
Here, $c_s$ is the sound speed evaluated at temperature $T_{eff}$, in order to take into account the backreaction of heating on structure formation. In this framework, the production of UV photons can be described as
\begin{equation}
\frac{dn_{\sm{ph}}/dz}{n_\fHI}\sim\xi \frac{df_{\sm{coll}}}{dz},
\end{equation}
where $\xi=A_{\fHeI}f_*f_{\sm{esc}}N_{\sm{ion}}$, with $A_{\fHeI}=4/(4-3Y_p)=1.22$, $N_{\sm{ion}}$ the number of ionizing photons per stellar baryon, $f_*$ is the star formation efficiency and $f_{\sm{esc}}$ the escape fraction of UV photons from their host galaxies. The quantity $f_{\sm{coll}}$ denotes the fraction of dark matter collapsed into halos, and is given as
\begin{equation}
f_{\sm{coll}}=\rm{erfc}\left[\frac{\delta_c(z)}{\sqrt{2}\sigma(M_{\sm{min}})} \right],\label{coll}
\end{equation}
where $M_{\sm{min}}=\mathrm{min}(M_F,10^5\ M_\odot)$, $\delta_c=1.69/D(z)$ is the linearized density threshold for collapse in the spherical top-hat model and $\sigma(M_{\sm{min}})$ describes the power associated with the mass scale $M_{\sm{min}}$.

{A relevant question in this context is also the role of Lyman-Werner (LW) feedback, which may suppress the star formation rate in low-mass halos. The role of such feedback has been addressed using different approaches. For instance, \citet{Machacek01}, \citet{OShea08} and \citet{WiseAbel07} have addressed this question employing numerical simulations in a cosmological context, assuming a constant LW-background radiation field. These simulations indicated that such feedback can delayed star formation considerably.} 

{More self-consistent simulations show, however, that the above calculations overestimated the role of LW-feedback. Considering single stellar sources and neglecting self-shielding, \citet{WiseAbel08} showed that LW-feedback only marginally delays star formation in halos that already started collapsing before the nearby star ignites. More detailed simulations taking into account self-shielding show that the star formation rate may be changed by only $20\%$ in the presence of such feedback \citep{Johnson07b}. This is due to the rapid re-formation of molecular hydrogen in relic HII regions, which leads to abundances of the order $10^{-4}$. Such abundances effectively shield against LW-feedback and make it ineffective \citep{Johnson08}. This is the point of view adopted here, which may translate into an uncertainty of $\sim20\%$ in the star formation rate. In fact, in scenarios involving dark matter annihilation, \HzI formation and self-shielding could be even further enhanced compared to the standard case \citep{Mapelli07}.
}

The models have to reproduce the reionization optical depth given by $\tau=0.087\pm0.017$ \citep{Komatsu08} and fully ionization at $z\sim6$ \citep{Becker01}. In the following, we will try to construct appropriate reionization histories for the different dark star models.

\subsection{Reionization with MS-dominated dark stars}
\bef
\showone{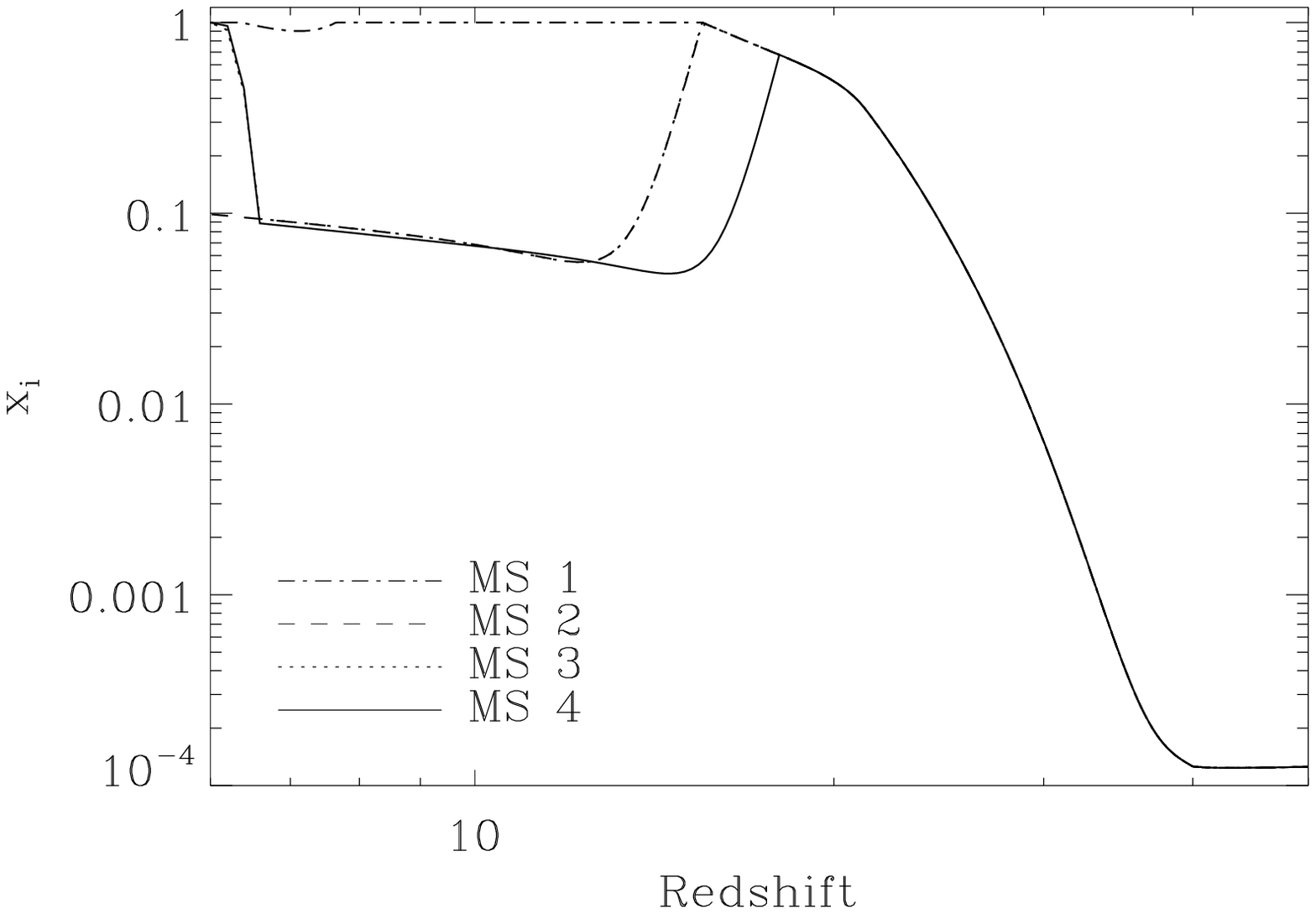}
\caption{The evolution of the effective ionized fraction $x_{\sm{eff}}$, for reionization models with main-sequence dominated dark stars (see Table \ref{tab:reionmsmodels}). Models MS 1 and MS 2 can be ruled out by reionization constraints, while models MS 3 and MS 4 require a sudden increase in the star formation rate by a factor of $30$ at redshift $6.5$. It appears more realistic to assume lower masses and star formation efficiencies to reconcile dark star models with observations.}
\label{ionms}
\eef
As shown previously \citep{SchleicherBanerjee08a}, MS-dominated dark stars with $\sim1000\ M_\odot$ would significantly overproduce the reionization optical depth if this type of stars had been common throughout the early universe. If, on the other hand, MS-dominated stars only had mass scales of $\sim100\ M_\odot$, comparable to conventional Pop.~III stars, reionization could not discriminate between them and conventional Pop.~III stars, and dark stars would be compatible with observations. Alternatively, as explained in the introduction, a transition in the stellar population might help to alleviate the problem for high-mass dark stars. We will explore this possibility in more detail to work out whether such a scenario is conceiveable.

Numerical simulations by \citet{Dove00}, \citet{Ciardi02} and \citet{Fujita03} indicated rather high escape fractions of order $100\%$ for massive Pop.~III stars. \citet{Wood00} found rather low escape fractions below $10\%$, while radiation hydrodynamics simulations by \citet{Whalen04} show that such stars can easily photo-evaporate the minihalo. Here we adopt the point of view that indeed massive stars can photoevaporate small minihalos, but that the escape fraction will be reduced to $\sim10\%$ in atomic cooling halos that have virial temperatures larger than $10^4$ K. Thus, we set $f_{\sm{esc}}=1$ if the filtering mass is below the mass scale $M_c=5\times10^7M_\odot \left(\frac{10}{1+z} \right)^{3/2}$ that corresponds to the virial temperature of $10^4$ K \citep{Oh02,Greif08}, and $f_{\sm{esc}}=0.1$ in the other case. To reflect the expected stellar mass of $\sim800\ M_\odot$, we choose a star formation efficiency of $f_*\sim1\%$, an order of magnitude higher than what we expect for conventional Pop.~III stars \citep{SchleicherBanerjee08a}.

 Assuming that reionization is completely due to these MS-dominated dark stars (model MS 1), we find that the universe is fully ionized at redshift $z_{\sm{reion}}=15.5$ and the reionization optical depth is $\tau_{\sm{reion}}\sim0.22$, i. e. significantly larger than the WMAP 5 optical depth (see Fig. \ref{ionms}). Such a model is clearly ruled out. 

To reconcile the presence of such massive dark stars with observations, one could invoke a double-reionization scenario, assuming a transition to a different mode of star formation induced by the strong UV feedback of MS-dominated dark stars. In fact, even for conventional star formation models, it is discussed that such UV feedback may lead to a less massive mode of star formation  \citep{Johnson06,Yoshida07a,Yoshida07b}. In addition, chemical enrichment should facilitate such a transition as well \citep{Schneider06,Clark08,Omukai08,Smith08a, Smith08b, GreifGlover08}, although it is unclear how well metals will mix with the pristine gas. We assume that the transition to a low-mass star formation mode with a Scalo-type IMF \citep{Scalo98} happens at redshift $15.5$, when the universe is fully ionized and UV feedback fully effective. For the subsequent Pop.~II stars, we assume a star formation efficiency of $f_*=5\times10^{-3}$ and $N_{\sm{ion}}=4\times10^3$ UV photons per stellar baryon.

Corresponding photon escape fractions are highly uncertain. Observations of \citet{Steidel01} indicate an escape fraction of $10\%$ at $z\sim3$, while others find detections or upper limits in the range $5-10\%$ \citep{Giallongo02,Malkan03,Fernandez03,Inoue05}. We adopt the generic value of $10\%$ for simplicity, though our results do not strongly depend on this assumption. For this scenario, to which we refer as model MS 2, we find an optical depth $\tau_{\sm{reion}}=0.082$ well within the WMAP constraint, but the universe does not get fully ionized until redshift zero. This scenario is thus rejected based on the constraint from quasar absorption spectra \citep{Becker01}.
\\ \\
To fulfill both the WMAP constraint as well as full-ionization at $z\sim6$, we need to introduce an additional transition in our model. At redshift $z_{\sm{burst}}=6.5$, we increase the star formation efficiency to $15\%$. This might be considered as a sudden star burst and results in full-ionization at $z=6.2$. In this case, we find $\tau_{\sm{reion}}=0.116$, which is within the $2\sigma$ range of the WMAP data. However, we are not aware of astrophysical models that provide a motivation for such a sudden star burst that increases the star formation rate by a factor of $30$. Based on gamma-ray burst studies, \citet{Yuksel08} showed that the cosmic star formation rate does not change abruptly in the redshift range between redshift zero and $z_{\sm{burst}}=6.5$. Such a sudden burst is thus at the edge of violating observation constraints.

To improve the agreement with WMAP, one can consider to shift the first transition to $z_{\sm{Pop II}}=18$ where full ionization is not yet reached (model MC 4), which yields an optical depth $\tau_{\sm{reion}}=0.086$, in good agreement with WMAP. At this redshift, $68\%$ of the universe are already ionized, so UV feedback might already be active and induce a transition in the stellar population. The results are given in Fig. \ref{ionms} and summarized in Table \ref{tab:reionmsmodels}.

\begin{table}[htdp]
\begin{center}
\begin{tabular}{ccccc}
Model & $z_{\sm{Pop II}}$ & $z_{\sm{burst}}$  & $\tau_{\sm{reion}}$ & $z_{f}$  \\
\hline
MS 1  &   -    &   -  & $0.22$  &  $15.5$  \\
MS 2  &   $15.8$ &   -  & $0.078$ & never  \\
MS 3  &   $15.5$ & $6.5$ & $0.116$ & $6.2$  \\
MS 4  &   $18.$  & $6.5$  & $0.086$ & $6.2$ \\
\hline
\end{tabular}
\end{center}
\caption{Reionization models for MS-dominated dark stars. The parameters $z_{\sm{Pop II}}$ and $z_{\sm{burst}}$ give the transition redshifts to a mode of Pop.~II star formation and to the sudden star burst, while $\tau_{\sm{reion}}$ is the calculated reionization optical depth and $z_{f}$ the redshift of full ionization. }
\label{tab:reionmsmodels}
\end{table}%
However, we find that only models MS 3 and MS 4 cannot be ruled out observationally. These models require two severe transitions in the stellar population and cannot be considered as "natural". Improved measurements of the reionization optical depth from Planck \footnote{http://www.rssd.esa.int/index.php?project=planck} will remove further uncertainties and may rule out model MS 3 as well. From a theoretical point of view, it must be checked whether strong UV feedback can lead to the required transition to a low-mass star population, and in addition, the plausibility of a sudden star burst near redshift $6$ must be examined as well. In summary, it seems more plausible to conclude that MS-dominated dark stars were less massive than suggested by \citet{FreeseSpolyar08}, as already hinted by \citet{SchleicherBanerjee08a}.

\subsection{Reionization with CD dark stars}\label{reionizationcd}
\bef
\showone{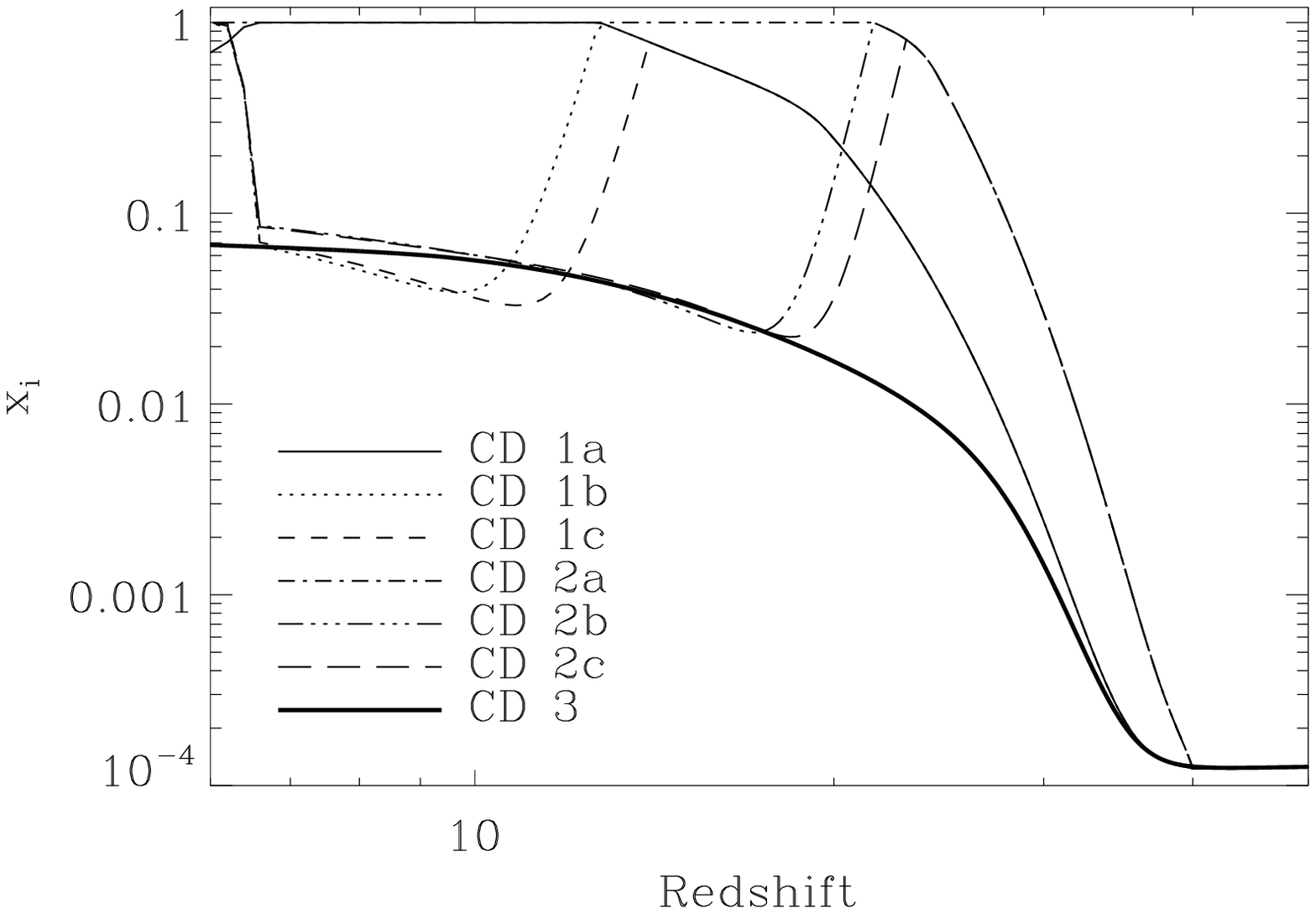}
\caption{The evolution of the effective ionized fraction $x_{\sm{eff}}$, for reionization models with capture-dominated dark stars (see Table \ref{tab:reioncdmodels}). Models CD 1a, CD 1b, CD 2a, CD 2b and CD 3 are ruled out due to reionization constraints, while the remaining models require an artificial star burst.}
\label{ioncd}
\eef

For CD dark star models, the situation is complicated by the fact that the number of UV photons per stellar baryon, $N_{\sm{ion}}$ , is model-dependent and changes with the  {environmental dark matter density}, $\rho_X$. We select three representative models of \citet{Yoon08}{, which assume a spin-dependent scattering cross section of $5\times10^{-39}$~cm$^2$ (see Table \ref{tab:reioncdmodels}). In general, stellar models depend on the product of this scattering cross section with the threshold dark matter density at the stellar radius \citep{Taoso08, Yoon08}. Lower elastic scattering cross sections therefore correspond to going to smaller threshold densities at the same elastic scattering cross section.}

 In the models CD 1 and 2, $N_{\sm{ion}}$ is larger than for conventional Pop.~III stars, while in the model CD 3, it is even less than in the case of Scalo-type Pop.~II stars. Such a low luminosity is unlikely to photo-evaporate star-forming halos, and we thus adopt $f_{\sm{esc}}=10\%$ for this case. However, such Scalo-type Pop.~II stars are ruled out as sole sources for reionization \citep{SchleicherBanerjee08a}. As we show in Fig. \ref{ioncd}, even with a high star formation efficiency of $f_*=1\%$, they never ionize the universe completely.

 In principle, one could consider the presence of other sources to ionize the universe. While dark stars of type CD 3 may be the first stars to form, one might envision a transition to a stellar population with the power to ionize the universe. This transition is unlikely due to UV feedback, as UV feedback from dark stars is rather weak in this scenario. One thus has to rely on effective mixing of the produced metallicity, or assume that the first stellar clusters in atomic cooling halos contain a sufficient number of massive stars to reionize the universe \citep{Clark08}.

For the other two models, $N_{\sm{ion}}$ is significantly larger and we adopt the procedure from the previous subsection, such that $f_{\sm{esc}}$ depends on the filtering mass. We adopt a star formation efficiency of $f_*=0.1\%$. We examine the reionization models given in Table \ref{tab:reioncdmodels}, which essentially follow the philosophy of the models from the previous section. We calculate the reionization history for the case where these dark stars are sole sources (CD 1a, CD2a) and find that the optical depth is considerably too high. We then determine the redshift where the universe is fully ionized and assume a transition to Pop.~II stars at this redshift. In addition, to obtain full ionization at redshift $6$, we assume a late star burst as in the models MS 3 and MS 4. This approach corresponds to the models CD 1b and CD 2b, and yields optical depth that are at least within the $2\sigma$ error of WMAP 5. In the models CD 1c and CD 2c, we improve the agreement with WMAP by introducing the Pop.~II transition at an earlier redshift.

\begin{table*}[htdp]
\begin{center}
\begin{tabular}{lrrrrr}
Reion. model & $\rho_X/10^{12}$ & $N_{ion}$ & $f_*$ & $z_{\sm{Pop II}}$  & $\tau_{\sm{reion}}$   \\
\hline
CD 1a  &  $0.01\ \mathrm{GeV\ cm}^{-3}$ & $1.75\times10^5$ &   $0.1\%$ & - & $0.162$    \\
CD 1b  &  $0.01\ \mathrm{GeV\ cm}^{-3}$ & $1.75\times10^5$ &  $0.1\%$  & $12.7$ & $0.109$   \\
CD 1c  &  $0.01\ \mathrm{GeV\ cm}^{-3}$ & $1.75\times10^5$ & $0.1\%$  & $14.5$ & $0.089$  \\
CD 2a  &  $0.05\ \mathrm{GeV\ cm}^{-3}$ & $2.4\times10^6$ &  $0.1\%$  & - & $0.283$ \\
CD 2b  &  $0.05\ \mathrm{GeV\ cm}^{-3}$ & $2.4\times10^6$ &  $0.1\%$  & $21.6$ & $0.106$  \\
CD 2c  &  $0.05\ \mathrm{GeV\ cm}^{-3}$ & $2.4\times10^6$ & $0.1\%$  & $23$ & $0.084$  \\
CD 3  &   $1\ \mathrm{GeV\ cm}^{-3}$    & $1.1\times10^3$ & $1\%$  & - & $0.004$  \\
\hline
\end{tabular}
\end{center}
\caption{Reionization models for CD dark stars stars. The number of ionizing photons was determined from the work of \citet{Yoon08}. The parameters $z_{\sm{Pop II}}$ and $z_{\sm{burst}}$ give the transition redshifts to a mode of Pop.~II star formation and to the sudden star burst, while $\tau_{\sm{reion}}$ is the calculated reionization optical depth and $z_{f}$ the redshift of full ionization. {The calculation assumes a spin-dependent scattering cross section of $5\times10^{-39}$~cm$^2$. As stellar models depend on the product of this cross section with the threshold dark matter density, the effect of a lower scattering cross section is equivalent to a smaller threshold density.}}
\label{tab:reioncdmodels}
\end{table*}%

 The results are given in Fig. \ref{ioncd}. Again, it turns out that somewhat artificial models are required to allow for an initial population of CD dark stars. The best way to reconcile these models with the constraints from reionization might be to focus on those models that predict a parameter $N_{\sm{ion}}$ which is closer to the Pop.~III value of $4\times10^4$. This may be possible, as the transition from the models CD 1 and 2 to CD 3 is likely continuous, and an appropriate range of parameters may exist to re-concile models with observations. This would require a $\rho_X$ between $10^{11} \ \mathrm{GeV\ cm}^{-3}$ and $10^{12}\ \mathrm{GeV\ cm}^{-3}$. {As mentioned earlier, the apparent violation of reionization constraints by some models depends also on the uncertainties in the stellar lifetime. If the dark matter reservoir near the star is destroyed earlier due to dynamical processes, the lifetime may be significantly reduced. Also, we stress that the conclusions depend on the adopted elastic scattering cross section and the dark matter density in the environment. The discussion here is limited to those models that have previously been worked out in detail.  }

\section{Predictions for 21 cm observations}\label{21cm}
\bef
\showthree{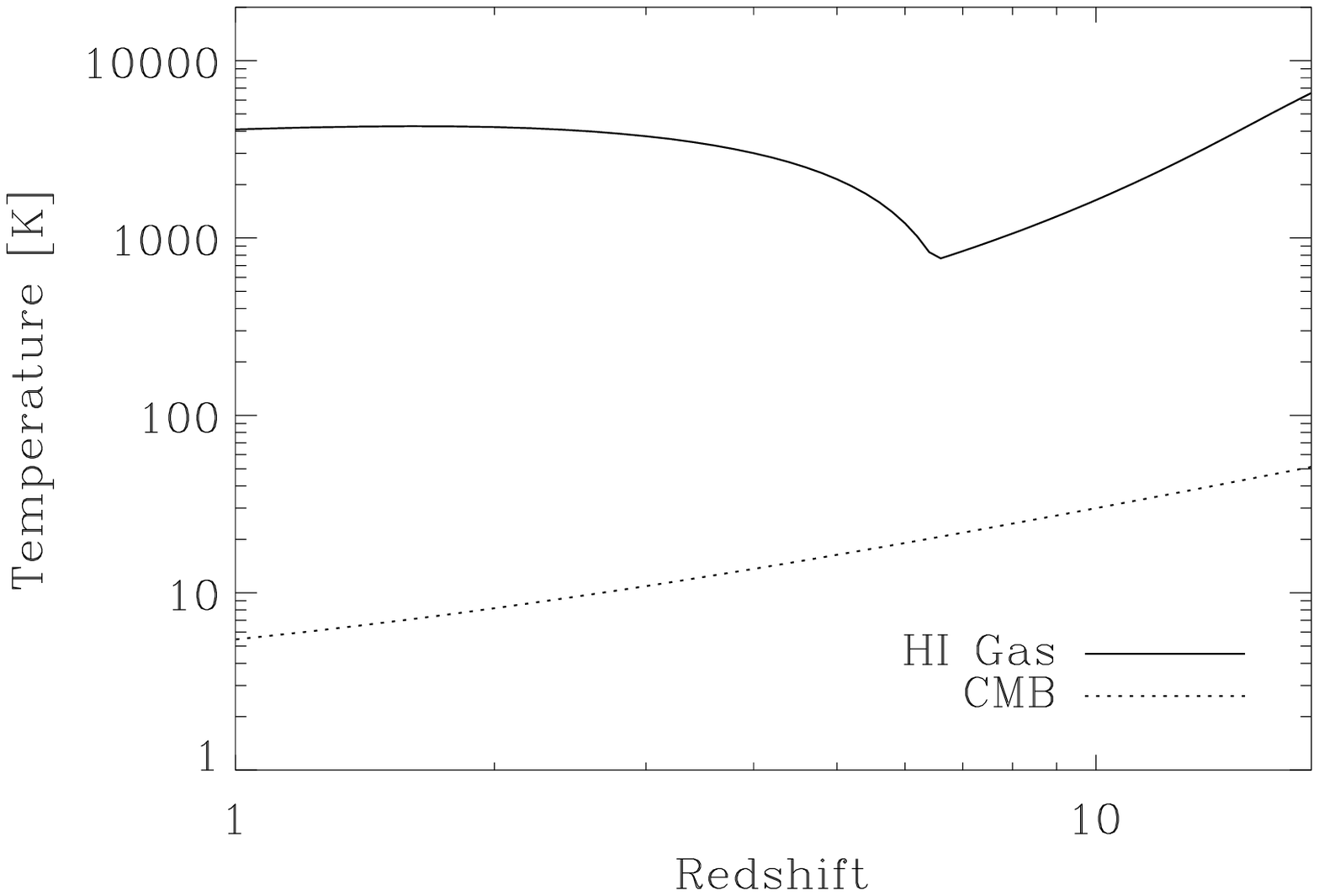}{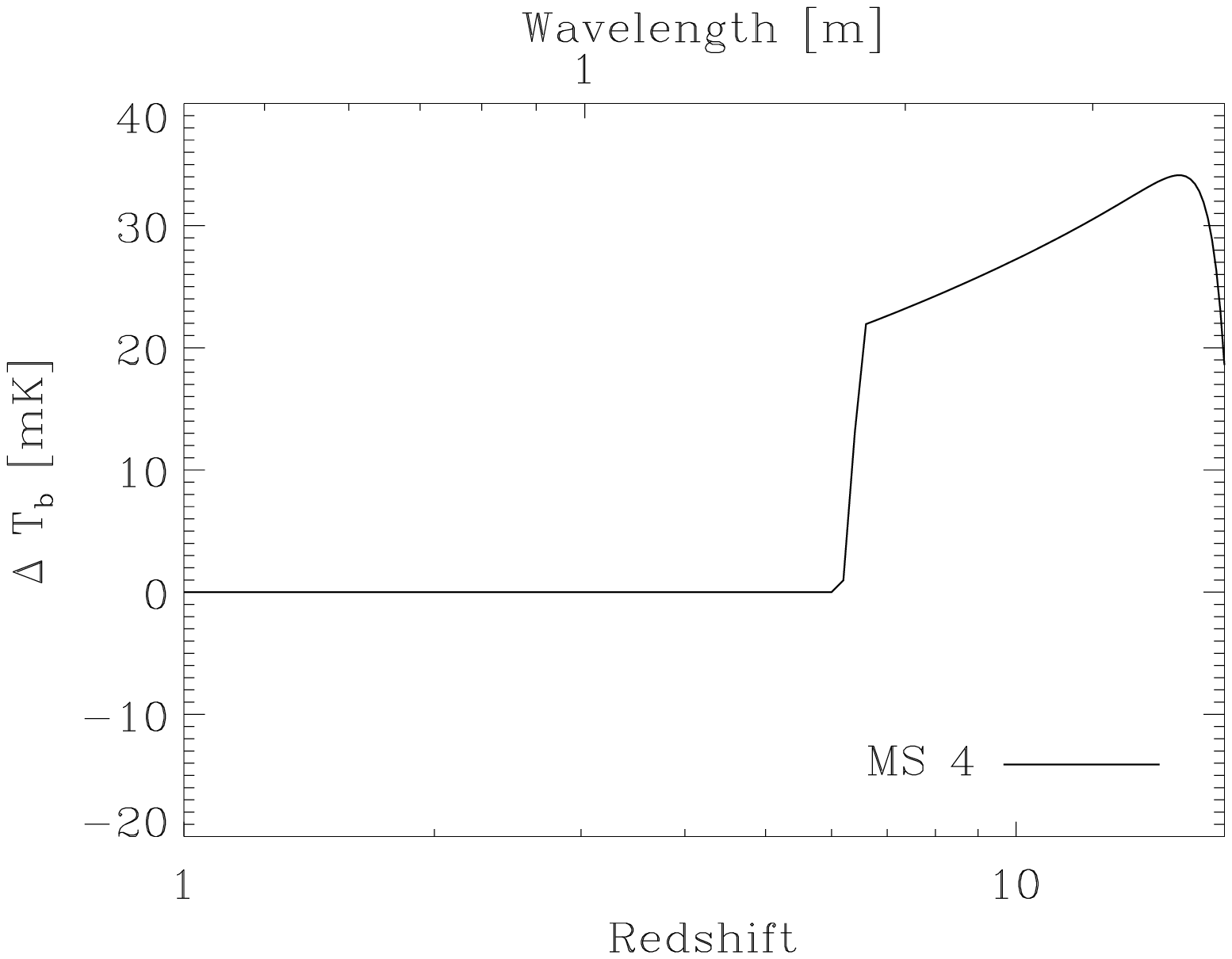}{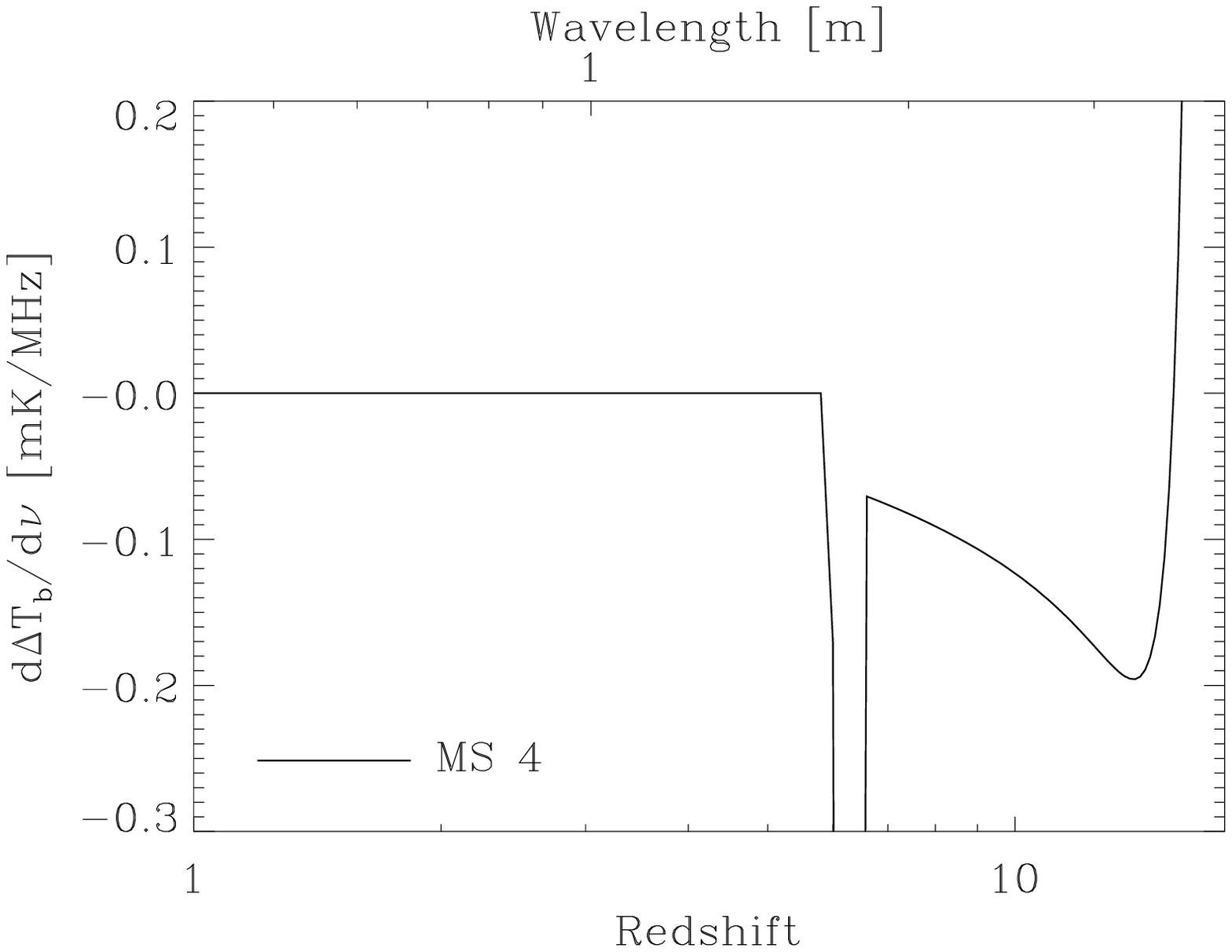}
\caption{$21$ cm signatures of double-reionization scenarios (here MS 4 from Table \ref{tab:reionmsmodels}). Given is the evolution after the first reionization phase, when the \HI gas is heated from the previous ionization. Top: HI gas temperature, here identical to the spin temperature. Middle: Expected mean $21$ cm brightness fluctuation. Bottom: Frequency gradient of the mean $21$ cm brightness fluctuation.}
\label{fig:21cm}
\eef
While some of the models suggested above essentially co-incide with standard reionization by mimicing the effects of conventional Pop.~III stars, others may have a very distinctive signature, as they consist of a double-reionization phase, and upcoming $21$ cm telescopes like LOFAR \footnote{http://www.lofar.org/} or SKA \footnote{http://www.skatelescope.org/} can thus verify or rule out such suggestions. The calculation shown in Fig. \ref{fig:21cm} is based on the double reionization model MS 4, but clearly the models MS 3, CD 1b, CD 1c, CD 2b and CD 2c yield similar results. In such a double-reionization scenario, the gas is heated to $\sim10^4$ K during the first reionization epoch. Assuming that the first reionization epoch ends at redshift $z_{\sm{Pop II}}$, the gas temperature in the non-ionized medium will then evolve adiabatically as 
\begin{equation}
T\sim10^4\ \mathrm{K}\left(\frac{1+z}{1+z_{\sm{Pop II}}} \right)^{2}.
\end{equation}
In addition, the previous reionization phase will have established a radiation continuum between the Lyman $\alpha$ line and the Lyman limit, where the universe is optically thin, apart from single resonances corresponding to the Lyman series. This radiation is now redshifted into the Lyman series and may couple the spin temperature $T_{\sm{spin}}$ of atomic hydrogen to the gas temperature T via the Wouthuysen-Field effect \citep{Wouthuysen52, Field58}. In fact, a small amount of Lyman $\alpha$ radiation suffices to set $T_{\sm{spin}}=T$ \citep{BarkanaLoeb05, Hirata06, Pritchard06}, which we assume here. Also, as the universe is optically thin to this radiation background, even Pop.~II sources will suffice to couple the spin temperure to the gas temperature. The mean $21$ cm brightness temperature fluctuation is then given as
\begin{eqnarray}
\delta T_b &=& 27 x_{\fHI}(1+\delta)\left(\frac{\Omega_b h^2}{0.023} \right) \left(\frac{0.15}{\Omega_m h^2}\frac{1+z}{10} \right)^{1/2}\nonumber\\
&\times& \left(\frac{T_S-T_r}{T_S}\right) \left(\frac{H(z)/(1+z)}{dv_{||}/dr_{||}} \right) \ \mathrm{mK},
\end{eqnarray}
where $x_{\fHI}$ denotes the neutral hydrogen fraction, $\delta$ the fractional overdensity, $\Omega_b$, $\Omega_m$ the cosmological density parameters for baryons and total matter, $h$ is related to the Hubble constant $H_0$ via $h=H_0/(100 \mathrm{km/s/Mpc})$, $T_r$ the radiation temperature and $dv_{||}/dr_{||}$ the gradient of the proper velocity along the line of sight, including the Hubble expansion. We further calculate the frequency gradient of the mean $21$ cm brightness temperature fluctuation to show its characteristic frequency dependence. In Fig. \ref{fig:21cm}, we show the evolution of the gas temperature, the mean $21$ cm brightness fluctuation and its frequency gradient for model MS 4.\\ \\
 As pointed out above, we expect similar results for other double-reionization models because of the characteristic adiabatic evolution of the gas and spin temperature. The decrease of the spin temperature with increasing redshift is a unique feature that is not present in other models that like dark matter decay \citep{Furlanetto06} or ambipolar diffusion heating from primordial magnetic fields \citep{Sethi05, TashiroSugiyama06, SchleicherBanerjee08b}, which may also increase the temperature during and before reionization.

\section{Cosmic constraints on massive dark matter candidates}\label{xray}

{In typical dark star models, it is assumed that massive dark matter candidates like neutralinos with masses of the order $100$~GeV annihilate into gamma-rays, electron-positron pairs and neutrinos \citep{Spolyar08, Iocco08, FreeseGondolo08, FreeseBodenheimer08, FreeseSpolyar08, IoccoBressan08, Yoon08}. {Similar to the constraint on high-redshift quasars from the X-ray background \citep{Dijkstra04, Salvaterra05, SchleicherSpaans08}, the gamma-ray and neutrino backgrounds allow to constrain the model for and the amount of dark matter annihilation.} As detailed predictions for the decay spectra are highly model-dependent, it is typically assumed that roughly $1/3$ of the energy goes into each annihilation channel. Constraints on such scenarios are available from the Galactic center and the extragalactic gamma-ray and neutrino backgrouns \citep{Ullio02, Beacom07, Yuksel07, Mack08}. In this section, we consider how such constraints are affected when the increase in the annihilation rate due to enhanced dark matter densities after the formation of dark stars is taken into account.}

\subsection{Gamma-ray constraints}
\bef
\showone{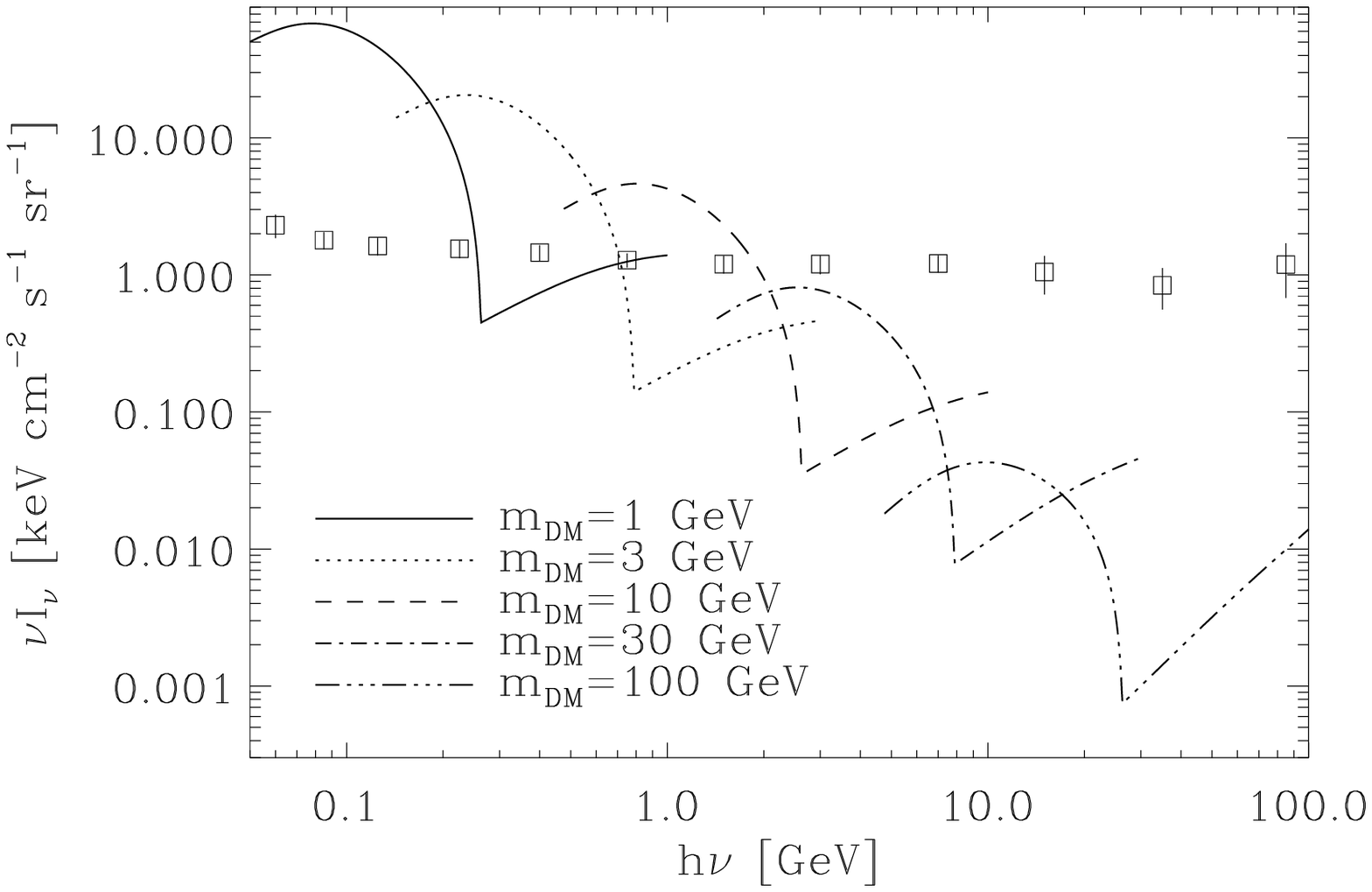}
\caption{The predicted gamma-ray background due to direct annihilation into gamma-rays in the presence of adiabatic contraction during the formation of dark stars, and the background measured by EGRET (squares) \citep{Strong04}. One finds two peaks in the annihilation background for a given particle mass: One corresponding to annihilation at redshift zero, and one corresponding to the redshift where the enhancement from adiabatic contraction was strongest. }
\label{fig:gamma}
\eef

{
We adopt the formalism of \citet{Mack08} who recently addressed the direct annihilation of massive dark matter particles into gamma-rays. The background intensity $I_\nu$ is given from an integration along the line of sight as
\begin{equation}
I_\nu=\frac{c}{4\pi}\int \frac{dz P_\nu([1+z]\nu,z)}{H(z)(1+z)^4},\label{intgamma}
\end{equation}
where $P_\nu(\nu,z)$ is the (proper) volume emissivity of gamma-ray photons, which is given as
\begin{equation}
P_{\nu}=\alpha_b\delta\left((1+z)\nu-m_{\sm{DM}} \right) \frac{m_{\sm{DM}}}{\mathrm{keV}}\ \mathrm{keV}\ \langle \sigma v \rangle n_{\sm{DM}}^2 C_{\gamma},\label{Pgamma}
\end{equation}
where $\langle \sigma v \rangle=3\times10^{-26}\ \mathrm{cm}^3\ \mathrm{s}^{-1}$ denotes the thermally-averaged annihilation cross section, $\alpha_b=1/3$ is the adopted branching-ratio to gamma-rays and $m_{\sm{DM}}$ the mass of the dark matter particle in keV. $C_{\gamma}$ refers to the dark matter clumping factor. This clumping factor depends on the adopted dark matter profile and the assumptions regarding substructure in a halo \citep{Ahn05c, Ando05, Chuzhoy08, Cumberbatch08}.  Here we use the clumping factor for a NFW dark matter profile \citep{Navarro97} which has been derived by \citet{Ahn05a, Ahn05c}. For $z<20$, it is given in the absence of adiabatic contraction as a power-law of the form
\begin{equation}
C_{\sm{DM}}=C_{\sm{DM}}(0)(1+z)^{-\beta},\label{clumpg}
\end{equation}
where $C_{\sm{DM}}(0)$ is the clumping factor at redshift zero and $\beta$ determines the slope. For a NFW profile \citep{Navarro97}, $C_{\sm{DM}}(0)\sim10^5$ and $\beta\sim1.8$. The enhancement due to adiabatic contraction is taken into account by defining
\begin{equation}
C_{\gamma}=C_{\sm{DM}}f_{\sm{enh}},
\end{equation}
where the factor $f_{\sm{enh}}$ describes the enhancement of the halo clumping factor due to adiabatic contraction (AC). We have estimated this effect based on the results of \citet{IoccoBressan08}, comparing a standard NFW profile with the enhanced profile that was created during dark star formation. We only compare them down to the radius of the dark star and find an enhancement of the order $\sim10^3$. For the NFW case, the clumping factor would be essentially unchanged when including smaller radii as well, while the AC profile is significantly steeper and the contribution from inside would dominate the contribution to the halo clumping factor. However, as the annihilation products are trapped inside the star, it is natural to introduce an inner cut-off at the stellar radius. In addition, we have to consider the range of halo masses and redshifts in which dark stars may form. We assume that the halo mass must be larger than the filtering mass to form dark stars. However, there is also an upper mass limit. Halos with masses above
\begin{equation}
M_c=5\times10^7M_\odot \left(\frac{10}{1+z} \right)^{3/2}
\end{equation}
correspond to virial temperatures of $10^4$ K \citep{Oh02} and are highly turbulent \citep{Greif08}. It seems thus unlikely that stars will form on the very cusp of the dark matter distribution in such halos, and more complex structures may arise. We thus assume that dark stars form in the mass range between $M_F$ and $M_c$. Once $M_c$ becomes larger than $M_F$, dark star formation must end naturally. In fact, it may even end before, as discussed in \S \ref{reionization}. To obtain the highest possible effect, we assume that dark stars form as long as possible. We thus have
\begin{equation}
f_{\sm{enh}}=\left(1+10^3\frac{f_{\sm{coll}}(M_F)-f_{\sm{coll}}(M_c)}{f_{\sm{coll}}(M_{\sm{F}})}\right).\label{factor}
\end{equation}
In Fig.~\ref{fig:gamma}, we compare the results with EGRET observations of the gamma-ray background \citep{Strong04}. In the absence of adiabatic contraction, the predicted background peaks at the contribution from redshift zero \citep{Mack08}. We find that the enhancement of annihilation due to adiabatic contraction produces a second peak in the predicted background which originates from higher redshifts. In this scenario, particle masses smaller than $30$~GeV can thus be ruled out.}

{
\subsection{Neutrino constraints}
\bef
\showone{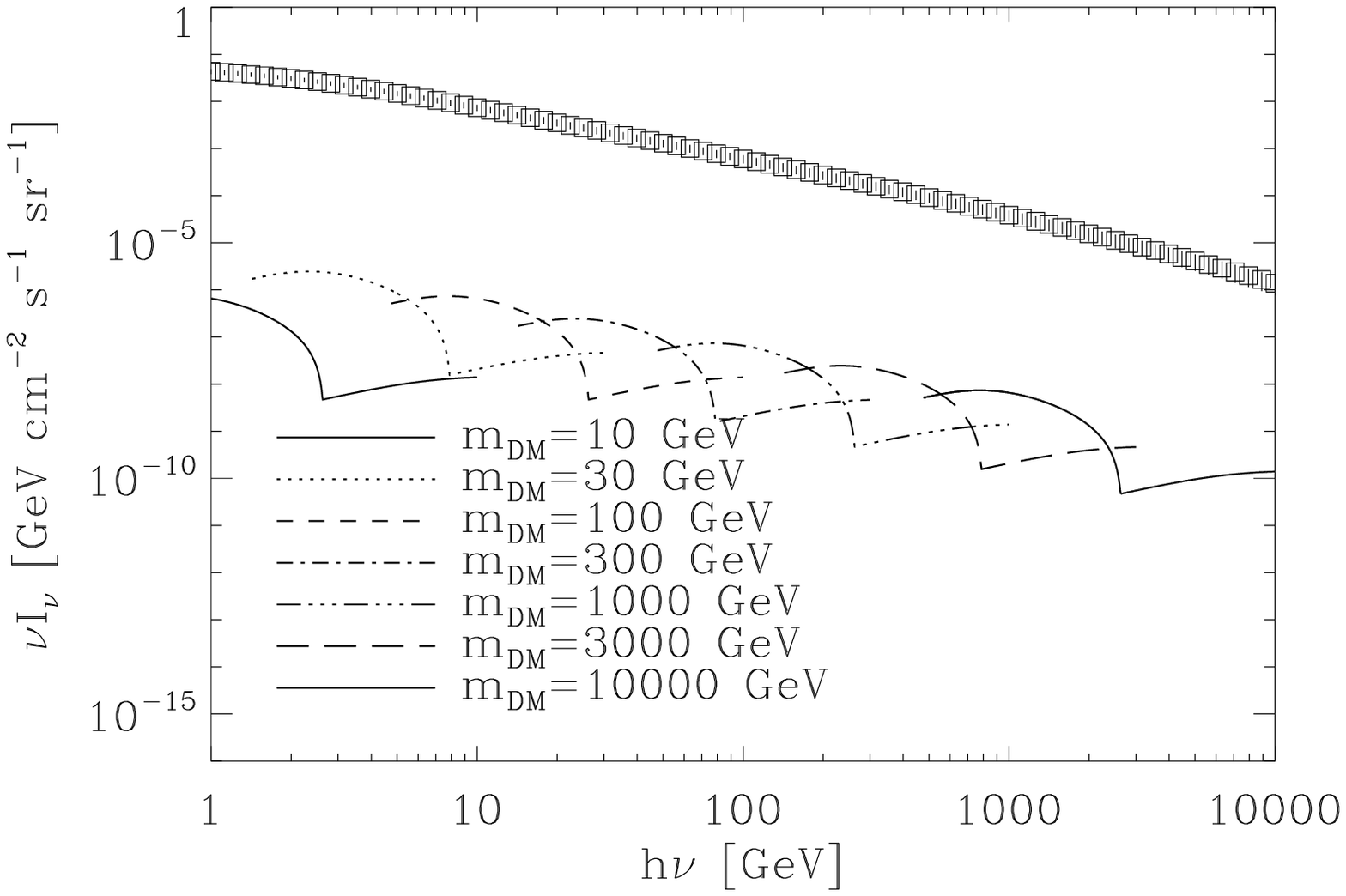}
\caption{The predicted neutrino background due to direct annihilation into neutrinos in the presence of adiabatic contraction during the formation of dark stars, and the atmospheric neutrino background \citep{Honda04}. One finds two peaks in the annihilation background for a given particle mass: One corresponding to annihilation at redshift zero, and one corresponding to the redshift where the enhancement from adiabatic contraction was strongest.}
\label{fig:neutrino}
\eef
The contribution to the cosmic neutrino flux can be obtained in analogy to Eq.~(\ref{intgamma}). As recent works \citep{Beacom07, Yuksel07}, we adopt an annihilation spectrum of the form
\begin{equation}
P_{\nu}=\alpha_b\delta\left((1+z)\nu-m_{\sm{DM}} \right) \frac{m_{\sm{DM}}}{\mathrm{keV}}\ \mathrm{keV}\ \langle \sigma v \rangle n_{\sm{DM}}^2 C_{\sm{neutrino}},
\end{equation}
which is analogous to the spectrum for annihilation into gamma-rays. The branching ratio to neutrinos is assumed to be $1/3$ as well, and the annihilation comes from the same dark matter distribution, thus yielding $C_{\sm{neutrino}}=C_{\gamma}$. The atmospheric neutrino background has been calculated from different experiments with generally good agreement \citep{Ahrens02, Gaisser02, Honda04, Ashie05, Achterberg07}. {\citet{Iocco08} adopted a similar atmospheric neutrino flux for comparison with the expected neutrino flux from dark stars.} We adopt here the data provided by \citet{Honda04} and compare them to the predicted background in Fig.~\ref{fig:neutrino}. The predicted background is always well below the observed background. }

{
\subsection{Emission from dark star remnants}\label{massiveremnant}
\bef
\showone{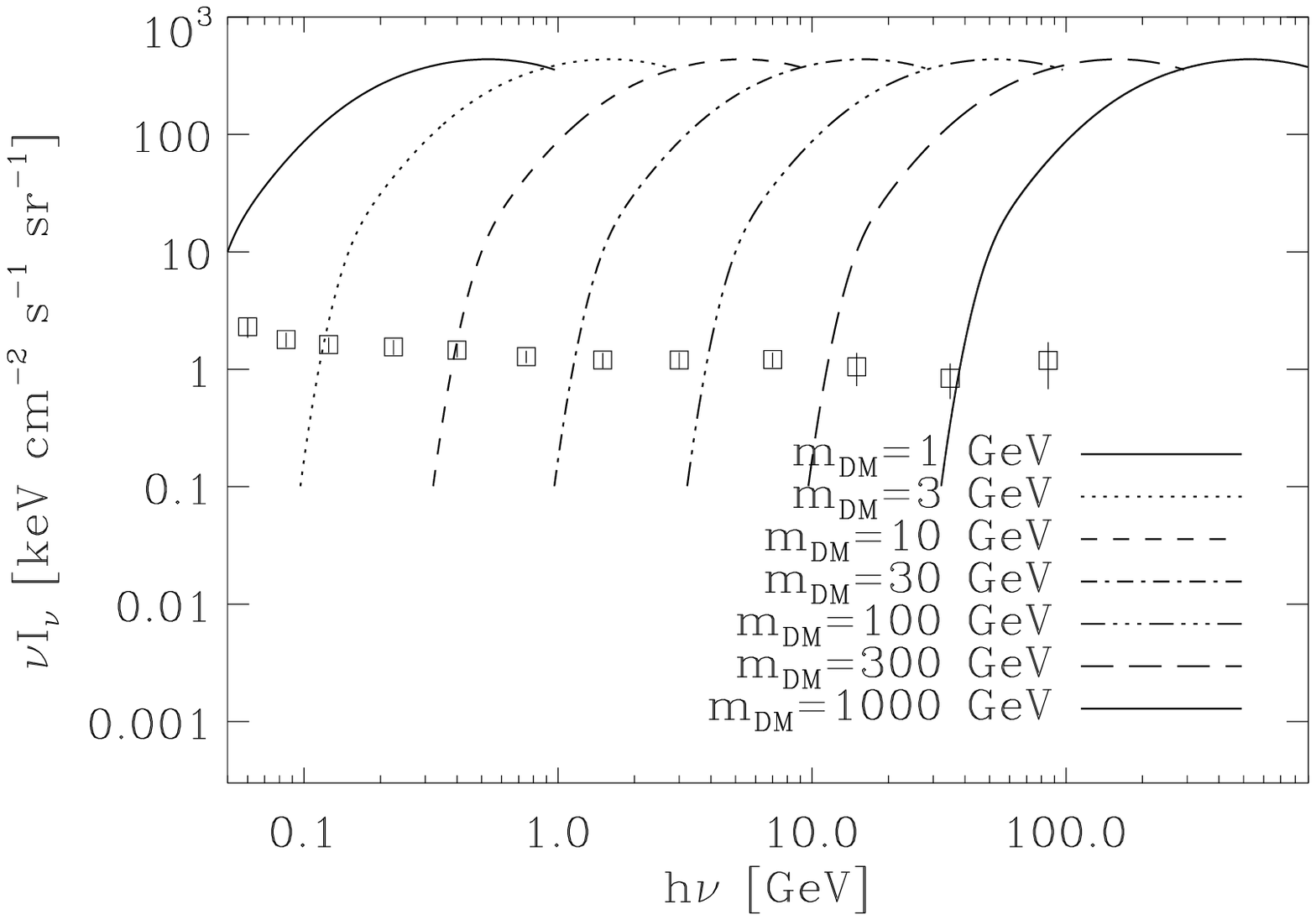}
\caption{The {\em maximum} gamma-ray background due to direct annihilation into gamma-rays in the remnants of dark stars, and the background measured by EGRET \citep{Strong04}. The actual contribution to the gamma-ray background is {highly model-dependent (see discussion in the text).}}
\label{fig:remnant}
\eef
In the previous subsections, we have included the enhancement of the halo clumping factor down to the stellar radius, as by definition the annihilation products on smaller scales are trapped inside the star. At the end of their lifetime, these stars may explode and the baryon density in the center may be largely depleted. {The dark matter density has certainly been significantly reduced due to annihilations during the lifetime of the star, but it may still be enhanced compared to the usual NFW case.} A detailed calculation of this effect is strongly model-dependent. As we have seen above, the strongest constraints are obtained for direct annihilation into gamma-rays, which is the case we pursue here in more detail.}

{
So far, we assumed that dark stars form in halos between the filtering mass $M_F$ and the mass corresponding to a virial temperature of $10^4$ K, $M_c$. To obtain an upper limit, it is sufficient to assume that in all halos above $M_F$ a dark star remnant will form at some point. Such an assumption clearly overestimates the total contribution at low redshift. When the dark star has formed, a fraction $f_{\sm{core}}\sim10^{-6}$ of the dark matter from the total halo is in the star \citep{FreeseBodenheimer08}. For the upper limit, we assume that the total amount of dark matter in star will contribute to the X-ray background (in fact, however, only the dark matter left over in the final remnant can contribute). In this case, we have a proper volume emissivity}
\begin{eqnarray}
P_{\nu}&=&\delta\left((1+z)\nu-m_{\sm{DM}} \right) \frac{m_{\sm{DM}}}{\mathrm{keV}}\ \mathrm{keV}\alpha_{511}\ f_{\sm{r}} f_{\sm{a}} \nonumber\\
&\times&n_{\sm{DM}}f_{\sm{core}}\frac{df_{\sm{coll}}(M_F)}{dt},
\end{eqnarray}{
where $n_{\sm{DM}}$ is the mean proper number density of dark matter particles, $m_{\sm{DM}}$ the particle mass in keV and $df_{\sm{coll}}/dt$ can be evaluated from Eq. (\ref{coll}). The model-dependent factor $f_r$ determines which fraction of the dark matter in the star will be left in the remnant. We adopt $f_{\sm{r}}=1$ to obtain an upper limit. The factor $f_{\sm{a}}$ determines the fraction of the remaining dark matter which actually annihilates, which we set to $f_{\sm{a}}=1$ as well. As in \S \ref{511}, $\alpha_{511}=1/4$ is the fraction of electron-positron annihilations per one dark matter annihilation process, corresponding to annihilation via positronium formation. In Fig. \ref{fig:remnant}, we compare the results with EGRET observations \citep{Strong04}. We find that the maximum contribution is clearly above the observed background.}

{Whether this maximum contribution can be reached, is however uncertain and the previous work in the literature only allows one to make rather crude estimates. For instance \citet{IoccoBressan08} calculate the density profile for a fiducial $100\,M_\odot$ protostar, finding that the density within the star roughly scales with $r^{-2}$ outside a plateau at a radius $r\sim10^{11}$~cm. At the stellar radius of $\sim10^{14}$~cm, the dark matter density is still $\sim10^{12}$~GeV~cm$^{-3}$. The timescale to remove this dark matter enhancement by annihilation is $\sim100$~Myr for $100$~GeV neutralinos. We need to estimate which fraction of the dark matter inside the star will be left at the end of its life, where the gas density is expelled by a supernova explosion and the dark matter annihilation from this region may contribute to the cosmic gamma-ray background.} 

{ \citet{Yoon08} adopted a timescale of $100$~Myr, the typical merger timescale at these redshifts, as the maximum lifetime for dark stars. 
Depending on the scattering cross section and the environmental density, the actual dark star lifetime may be considerably shorter. Indeed, as we showed in \S~\ref{reionizationcd}, it is difficult to reconcile lifetimes of $\sim100$~Myr with appropriate reionization scenarios. It is therefore reasonable to assume shorter timescales. In such a case, a reasonable estimate is that $\sim40\%$ of the dark matter inside the star would be left at the end of its life. This would still be enhanced compared to the standard NFW profile. In this case, the parameter $f_r$ is $\sim40\%$, and $f_a$ may be of order $1$, as the annihilation timescale is comparable to the Hubble time. We note that these numbers are highly uncertain, in particular regarding the exact evolution of dark matter density during the lifetime of the star, the effect of a supernova explosion on the dark matter cusp as well as the consequences of minor mergers. }


{There is however also a viable possibility that the dark matter distribution inside the star is significantly steeper than assumed above. In the case of dark matter capture by off-scattering from baryons, the dark matter density inside the star follows a Gaussian shape and is highly concentrated in a small region of $r\sim2\times10^9$~cm \citep{Griest87, Iocco08, Taoso08, Yoon08}. The implications are not entirely clear. If capture of dark matter stops at the end of the life of the star, the density inside the star will annihilate away quickly, and no significant contribution may come from the remnant. If, on the other hand, dark matter capture goes on until the end of the life of the star, a contribution to the background seems viable. In summary, this may provide a potential contribution to the cosmic background, but its strength is still highly uncertain and should be explored further by future work.} 

\subsection{Dependence of dark star models on the neutralino mass}
{We conclude this section with a discussion on the constraints from cosmic backgrounds for different neutralino masses. As dark star models in the literature mostly consider neutralinos of $100$~GeV, there are uncertainties that need to be addressed when considering different neutralino masses. For models involving the capture of dark matter, \citet{Iocco08} states that the mass of the neutralino does not change the annihilation luminosity. {\citet{Taoso08}} find that variations due to different neutralino masses are less than $5\%$. While these results may hold for high masses, \citet{Spergel85} showed that for neutralino masses below $4$~GeV, they would evaporate from the star, as scattering with baryons can upscatter them as well. }

{In addition, the AC phase may be modified as well, as the dark matter annihilation rate in this phase is degenerate in the parameter $\langle \sigma v\rangle / m_{\sm{DM}}$. \citet{IoccoBressan08} find that the duration of the AC phase may change by almost $50\%$ if the dark matter mass is changed by a factor of $2$. The effect of different neutralino masses is therefore uncertain and should be explored in more detail. We will however assume that the general behaviour involving adiabatic contraction in the minihalo is still similar, such that the calculations below are approximately correct also for different neutralino masses.}

\section{Cosmic constraints on light dark matter}\label{light}
{Observations of $511$ keV emission in the center of our Galaxy {\citep{Knodlseder03} provide recent motivation} to models of light dark matter \citep{Boehm04b}.} Such observational signatures can be explained assuming dark matter annihilation, while other models still have difficulties reproducing the observations \citep{BoehmHooper04}. The model assumes that dark matter annihilates into electron-positron pairs, which in turn annihilate into $511$ keV photons. Direct annihilation of dark matter into gamma-rays or neutrinos is assumed to be suppressed to avoid the gamma-ray constraints and to ensure a sufficient positron production rate. It is known that electron-positron annihilation occurs mainly via positronium-formation in our galaxy \citep{Kinzer01}. In addition, it was shown \citep{Beacom05} that dark matter annihilation to electron-positron pairs must be accompanied by a continuous radiation known as internal bremsstrahlung, arising from electromagnetic radiative corrections to the dark matter annihilation process.

Motivated by these results, it was proposed that internal bremsstrahlung from dark matter annihilation may be responsible for the gamma-ray background at energies of $1$-$20$ MeV \citep{Ahn05a}. Conventional astrophysical sources cannot explain the observed gamma-ray background at these frequencies \citep{Ahn05b}. A comparison of the observed and predicted background below $511$ keV yields constraints on the dark matter particle mass \citep{Ahn05c}. Here we examine whether and how this scenario is affected if dark stars form in the early universe. We use a thermally averaged cross section $\langle\sigma v\rangle\sim3\times10^{-26}\ \mathrm{cm}^3\ \mathrm{s}^{-1}$ to account for the observed dark matter density \citep{Drees93}. This implies that $\langle \sigma v\rangle$ is velocity-independent (S-wave annihilation). While \citet{BoehmHooper04} argue that S-wave annihilation overpredicts the flux from the galactic center, others argue that it is still consistent \citep{Ahn05a, Ahn05c}. The cross-section adopted here is well-within the conservative constraints of \citet{Mack08}. {The effect of light dark matter annihilation on structure formation in the early universe has been studied in various works, e. g. \citep{Mapelli06, Ripamonti07a, Ripamonti07b}. Constraints from upcoming $21$ cm observations have been explored by \citet{Furlanetto06} and \citet{Valdes07}, while constraints from background radiation have been considered by \citet{Mapelli05}.} {The effects of early dark matter halos on reionization have been addressed recently by \citet{Natarajan08}.}

{As in the previous section, we point out that significant uncertainties are present when considering dark star models for different dark matter masses, as this question is largely unexplored. In particular, we emphasize that no capturing phase will be present for light dark matter, as shown in the work of \citet{Spergel85}. Another uncertainty is the question whether to adopt self-annihilating dark matter (i. e. Majorana particles) or particles and antiparticles of dark matter. In the calculations below, we assume that light dark matter is self-annihilating. Otherwise, our results would be changed by a factor of $0.5$.}

\subsection{$511$ keV emission}\label{511}
\bef
\showone{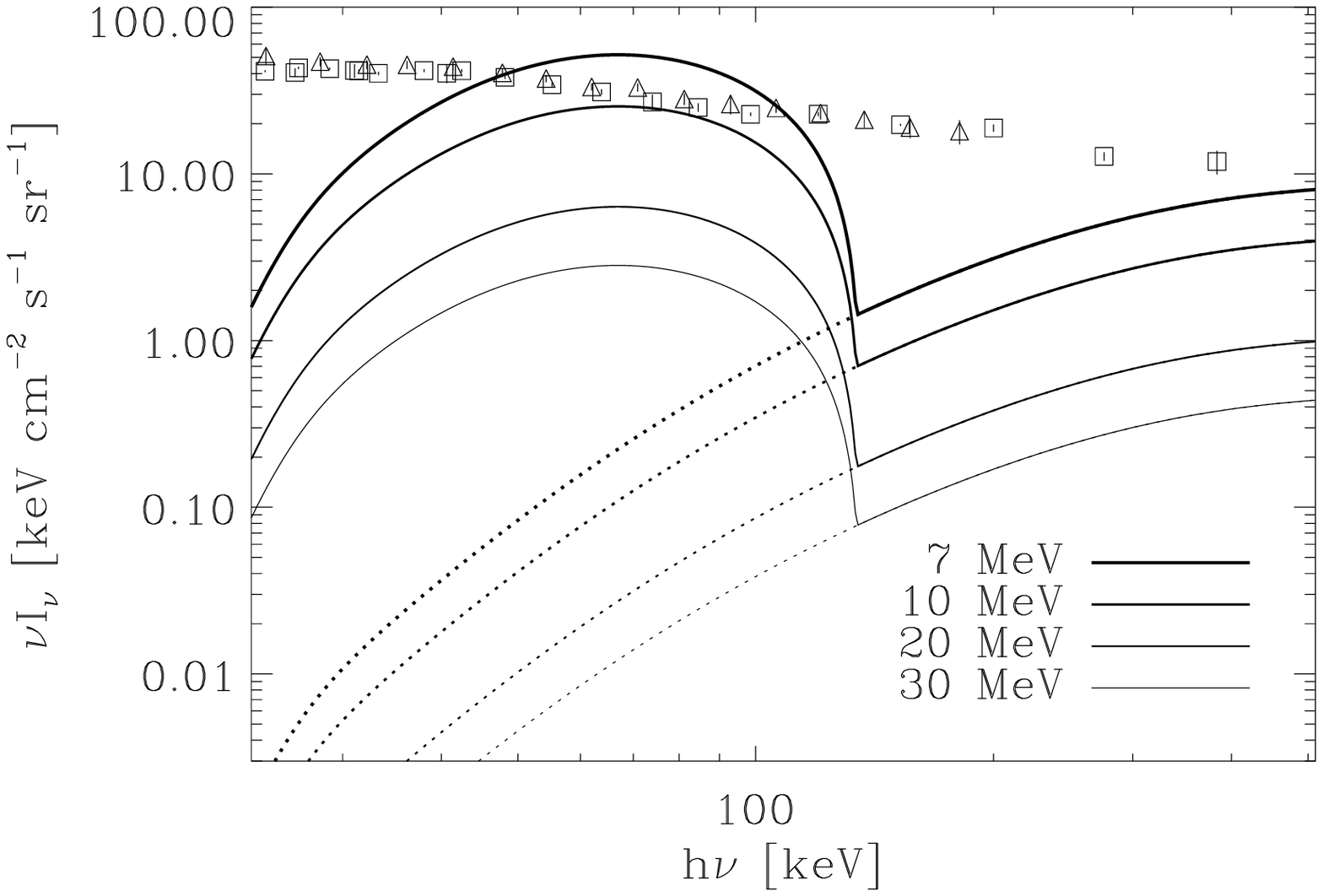}
\caption{The predicted X-ray background due to $511$ keV emission for different dark matter particle masses. Solid lines: Enhanced signal due from adiabatic contraction, dotted lines: Conventional NFW profiles. The observed X-ray background from the HEAO experiments (squares) \citep{Gruber99} and Swift/BATSE (triangles) \citep{Ajello08} is shown as well. The comparison yields a lower limit of $10$ MeV on the dark matter mass for the adiabatically contracted profiles, and $7$ MeV for standard NFW halo profiles.}
\label{fig:bg511}
\eef
{The expected X-ray background from $511$ keV emission is calculated from Eq.~(\ref{intgamma}). The volume emissivity of $511$ keV photons is given as}
\begin{equation}
P_{\nu}=\delta\left((1+z)\nu-\nu_{511} \right) 511\ \mathrm{keV}\alpha_{511}\ \langle \sigma v \rangle n_{\sm{DM}}^2 C_{511},
\end{equation}
where $\langle \sigma v \rangle$ denotes the thermally-averaged annihilation cross section, $\alpha_{511}$ is the fraction producing an electron-positron pair per dark matter annihilation process and $\nu_{511}$ the frequency corresponding to $511$ keV. In our galaxy, this process happens via positronium formation \citep{Kinzer01}, and we assume that the same is true for other galaxies. In $25\%$ of the cases, positronium forms in a singlet (para) state which decays to two $511$ keV photons, whereas $75\%$ form in a triplet (ortho) state which decays into a continuum. We thus adopt $\alpha_{511}=1/4$ for $511$ keV emission. $C_{511}$ refers to the dark matter clumping factor, which is still highly uncertain. The main uncertainty is due to the adopted dark matter profile and the assumptions regarding substructure in a halo \citep{Ahn05c, Ando05, Chuzhoy08, Cumberbatch08}.

 Here we use the clumping factor for a NFW dark matter profile \citep{Navarro97} which has been derived by \citet{Ahn05a, Ahn05c}, as to our knowledge, no calculations of dark star formation are available for other dark matter profiles. For $z<20$, it is given as a power-law as
\begin{equation}
C_{\sm{DM}}=C_{\sm{DM}}(0)(1+z)^{-\beta},\label{clump}
\end{equation}
where $C_{\sm{DM}}(0)$ describes the clumping factor at redshift zero and $\beta$ determines the slope. For a NFW profile \citep{Navarro97}, $C_{\sm{DM}}(0)\sim10^5$ and $\beta\sim1.8$. The effects of different clumping factors will be explored in future work \citep{SchleicherGlover08}.
\citet{Ahn05a,Ahn05c} included contributions from all halos with masses above a minimal mass scale $M_{\sm{min}}$, which was given as the maximum of the dark matter Jeans mass and the free-streaming mass. This approach assumes instantaneous annihilation of the created electron-positron pairs. As pointed out by \citet{Rasera06}, the assumption of instantaneuos annihilation is only valid if the dark matter halo hosts enough baryons to provide a sufficiently high annihilation probability, postulating this to happen in halos with more than $10^7-10^{10}\ M_\odot$, corresponding to their calculation of the filtering mass. We also calculate the filtering mass according to the approach of \citet{Gnedin98,Gnedin00}, but obtain somewhat lower masses, with $\sim10^5\ M_\odot$ halos at the beginning of reionization and $\sim3\times10^7 M_\odot$ at the end \citep{SchleicherBanerjee08a}. This is also in agreement with numerical simulations of \citet{Greif08} which find efficient gas collapse in halos of $10^5\ M_\odot$. The discrepancy may also be due to their different reionization model, which assumes reionization to start at redshift $20$.

We recall that the clumping factor can be considered as the product of the mean halo overdensity, the fraction of collapsed halos above a critical scale and the mean "halo clumping factor" that describes dark matter clumpiness within a halo. To take into account that electron-positron annihilation occurs only in halos above the filtering mass $M_{\sm{F}}$, we thus rescale the results of \citet{Ahn05a,Ahn05c} as
\begin{equation}
C_{511}=\frac{f_{\sm{coll}}(M_F)}{f_{\sm{coll}}(M_{\sm{min}})}C_{\sm{DM}}f_{\sm{enh}},
\end{equation}
where the factor $f_{\sm{enh}}$ {is given from Eq.~(\ref{factor}). As above, we assume that the halo mass must be larger than the filtering mass, and lower than the critical mass scale $M_c$ that corresponds to virial temperatures of $10^4$ K. For comparison,} we will also calculate $511$ keV emission with $f_{\sm{enh}}=1$. We note that the resulting background will be somewhat lower than the result of \citet{Ahn05c}, as we adopted $\alpha_{511}=1/4$ and {include only halos above the filtering mass scale in the} clumping factor. In Fig. \ref{fig:bg511}, we compare the results with the observed X-ray background from the HEAO-experiments \footnote{http://heasarc.gsfc.nasa.gov/docs/heao1/heao1.html} \citep{Gruber99} and SWIFT \footnote{http://heasarc.nasa.gov/docs/swift/swiftsc.html}/BATSE \footnote{http://www.batse.msfc.nasa.gov/batse/} observations \citep{Ajello08}. In the standard NFW case, we find a lower limit for the dark matter particle mass of $7$ MeV. For the case with adiabatically contracted profiles due to dark star formation, we find a slightly higher lower limit of $10$ MeV. This is because the enhancement is effective only for frequencies $h\nu<100$~keV, where the observed background is significantly larger than at $511$ keV, where \citet{Ahn05c} obtained their upper limit. 

\subsection{Internal Bremsstrahlung}
\bef
\showone{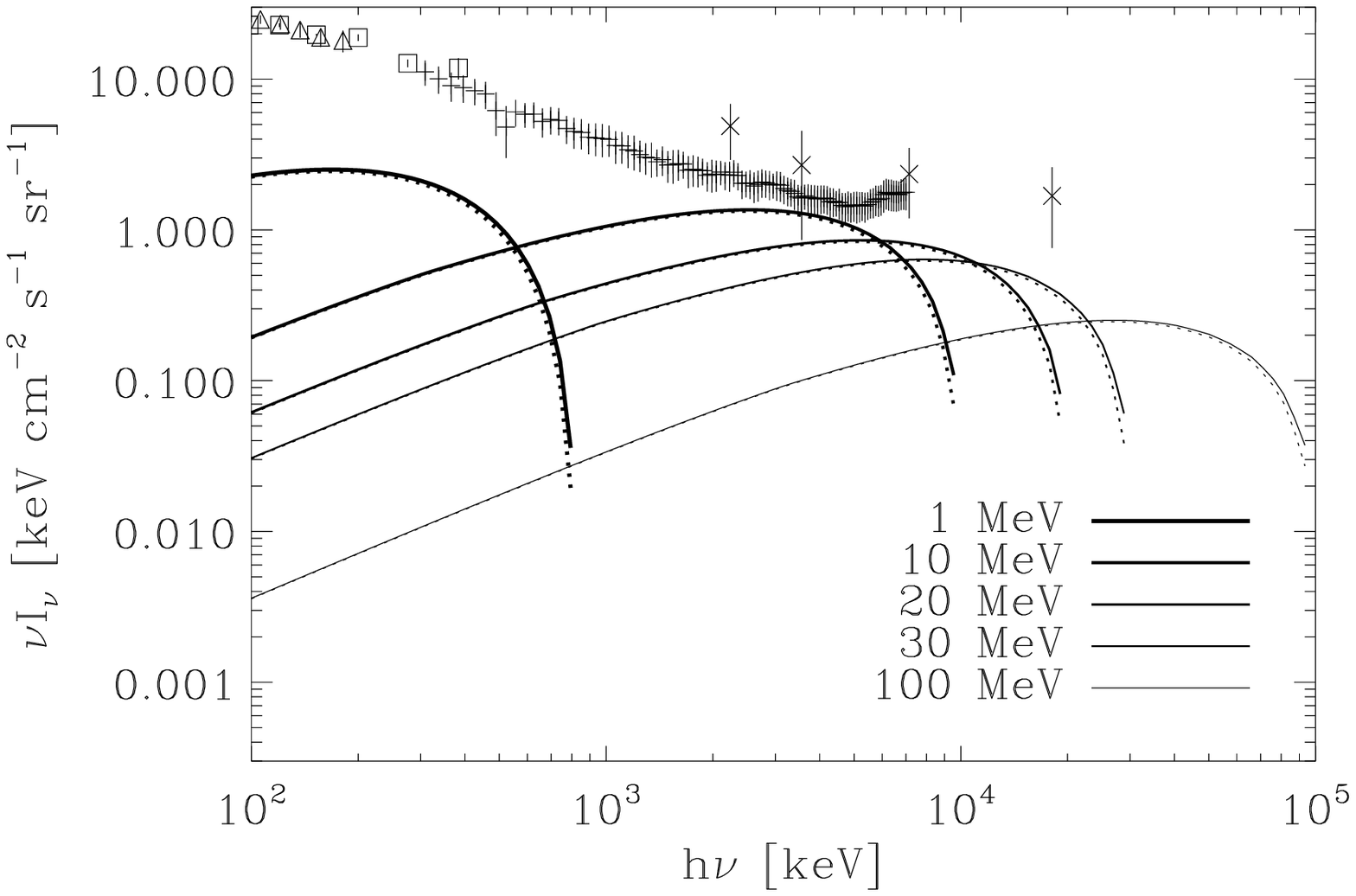}
\caption{The predicted gamma-ray background due to bremsstrahlung emission for different dark matter particle masses. Solid lines: Enhanced signal due from adiabatic contraction, dotted lines: Conventional NFW profiles. The lines overlap almost identically, as the main contribution comes from redshift zero, where the clumping factor is large and dark stars are assumed not to form. The observed gamma-ray background from the HEAO experiments (squares) \citep{Gruber99}, Swift/BATSE (triangles) \citep{Ajello08}, COMPTEL (crosses) \citep{Kappadath96} and SMM (plusses) \citep{Watanabe99} is shown as well.}
\label{fig:bgbrems}
\eef
The internal bremsstrahlung is calculated according to the approach of \citet{Ahn05a}. The background intensity is given by Eq. (\ref{intgamma}), with a proper volume emissivity 
\begin{equation}
P_\nu=\frac{1}{2}h\nu\langle\sigma v\rangle C_{\sm{brems}} n_{\sm{DM}}^2 \left[\frac{4\alpha}{\pi}\frac{g(\nu)}{\nu} \right],
\end{equation}
where $\alpha\sim1/137$ is the finestructure constant and $g(\nu)$ is a dimensionless spectral function, defined as
\begin{equation}
g(\nu)=\frac{1}{4}\left(\ln\frac{\tilde{s}}{m_e^2}-1 \right)\left[1+\left( \frac{\tilde{s}}{4m_{\sm{DM}}^2} \right)^2 \right],
\end{equation}
with $\tilde{s}=4m_{\sm{DM}}(m_{\sm{DM}}-h\nu)$. As \citet{Ahn05a} pointed out in a 'Note added in proof', bremsstrahlung is emitted in all dark matter halos, regardless of the baryonic content. There is thus no need to consider any shift in the minimal mass scale, the only thing to take into account is the enhancement of annihilation due to the AC profiles. The clumping factor $C_{\sm{brems}}$ is thus given as
\begin{equation}
C_{\sm{brems}}=C_{\sm{DM}}f_{\sm{enh}},
\end{equation}
where $f_{\sm{enh}}$ is given by Eq.~(\ref{factor}). In Fig. \ref{fig:bgbrems}, we compare the results with the observed gamma-ray background from the HEAO-experiments \citep{Gruber99} and SWIFT/BATSE observations \citep{Ajello08}, as well as SMM \footnote{http://heasarc.gsfc.nasa.gov/docs/heasarc/missions/solarmax.html} \citep{Watanabe99} and Comptel \footnote{http://wwwgro.unh.edu/comptel/} data \citep{Kappadath96}. We find that the signal is almost unchanged in the model taking into account dark star formation. The reason is that dark stars form mainly at high redshifts, in the range where $M_F<M_c$, while the dominant contribution to the background comes from redshift zero. Our results agree with \citet{Ahn05a}.

\subsection{Emission from dark star remnants}\label{remnants}
\bef
\showone{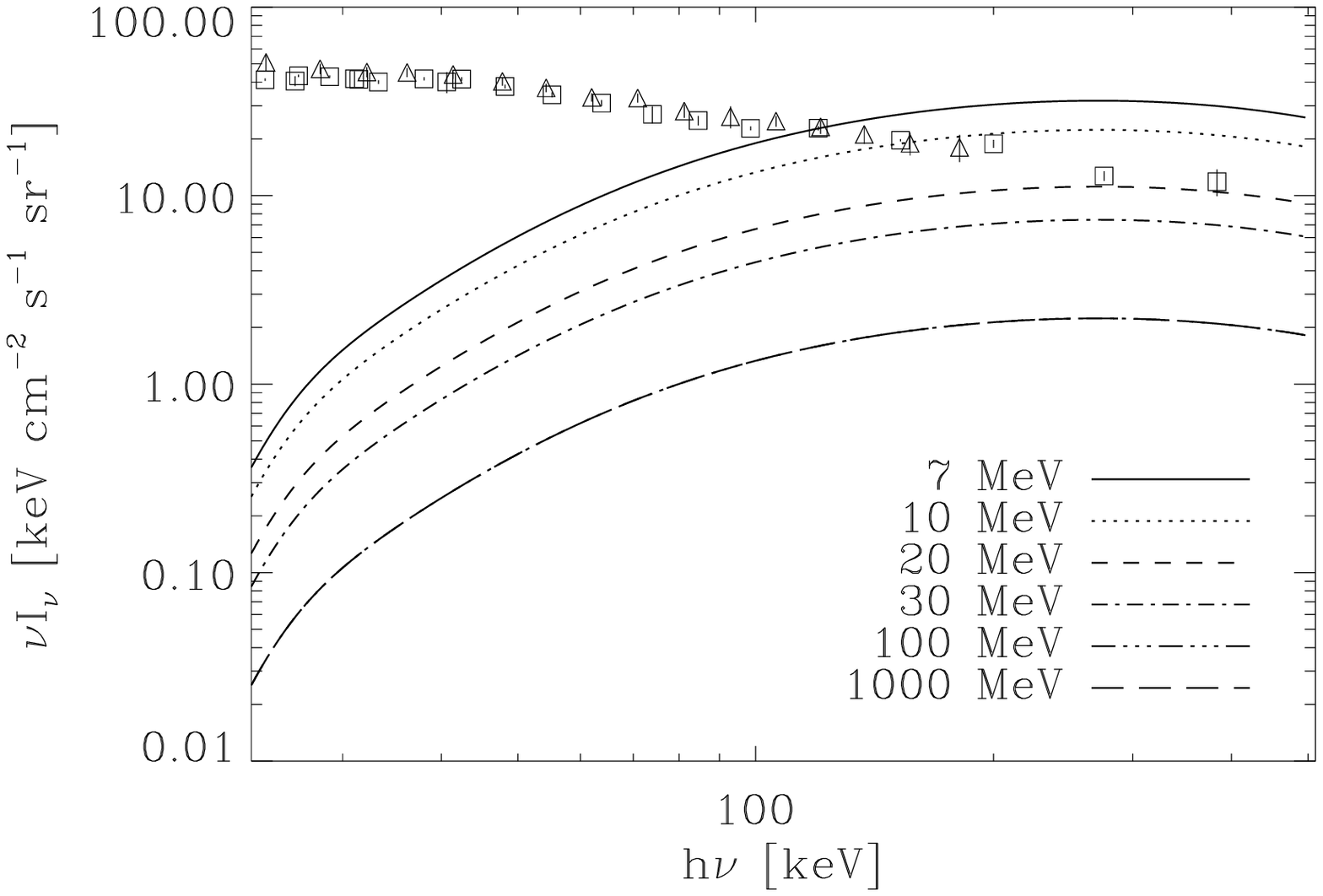}
\caption{The upper limit of X-ray radiation due to dark star remnants. The observed X-ray background from the HEAO experiments (squares) \citep{Gruber99} and Swift/BATSE (triangles) \citep{Ajello08} is shown as well. Only for very low dark matter particle masses, the upper limit is somewhat higher than the observed background. However, the actual contribution may be lower by some orders of magnitude (see discussion in the text).}
\label{fig:bgremnant}
\eef

{As in \S \ref{massiveremnant}, we consider a scenario where dark stars explode at the end of their lifetime and dark matter annihilation products in their remnants contribute to the cosmic background. For simplicity, we consider $511$ keV emission only, which is also mostly sensitive to modifications at high redshift. Again, we assume that dark stars form in halos between the filtering mass $M_F$ and the mass corresponding to a virial temperature of $10^4$ K, $M_c$. In this case, the volume emissivity is given as}
\begin{eqnarray}
P_{\nu}&=&\delta\left((1+z)\nu-\nu_{511} \right) 511\ \mathrm{keV}\alpha_{511}\ f_{\sm{r}} f_{\sm{a}} \nonumber\\
&\times&n_{\sm{DM}}f_{\sm{core}}\frac{df_{\sm{coll}}(M_F)}{dt},
\end{eqnarray}
where $n_{\sm{DM}}$ is the mean proper number density of dark matter particles and $df_{\sm{coll}}/dt$ can be evaluated from Eq. (\ref{coll}). The model-dependent factor $f_r$ determines which fraction of the dark matter in the star will be left in the remnant, we adopt $f_{\sm{r}}=1$ to obtain an upper limit. The factor $f_{\sm{a}}$ determines the fraction of the remaining dark matter which actually annihilates, which we set to $f_{\sm{a}}=1$ as well. As in \S \ref{511}, $\alpha_{511}=1/4$ is the fraction of electron-positron annihilations per one dark matter annihilation process, corresponding to annihilation via positronium formation. In Fig. \ref{fig:bgremnant}, we compare the results with the observed X-ray background from the HEAO-experiments \citep{Gruber99} and SWIFT/BATSE observations \citep{Ajello08}.

For dark matter particle masses below $30$ MeV, the upper limit found here is higher than the observed background. {Again, as discussed in \S~\ref{massiveremnant}, there are significant uncertainties regarding the question whether this high contribution can be reached, both due to uncertainties in the dark star models, which have not been explored for light dark matter, as well as the impact of a supernova explosion on the dark matter cusp. These possibilities should be addressed further in future work.}

\section{Summary and discussion}\label{outlook}
In this work, we have examined whether the suggestion of dark star formation in the early universe is consistent with currently available observations. We use these observations to obtain constraints on dark star models and dark matter properties. From considering cosmic reionization, we obtain the following results:
\begin{itemize}
\item Dark stars with masses of the order $800\ M_\odot$ as suggested by \citet{FreeseBodenheimer08} can only be reconciled with observations if somewhat artificial double-reionization scenarios are constructed. They consist of a phase of dark star formation followed by a phase of weak Pop.~II star formation and a final star burst to reionize the universe until redshift $6$. 
\item The same is true for dark stars in which the number of UV photons is significantly increased due to dark matter capture, as suggested by \citet{IoccoBressan08}.
\item It appears more reasonable to require that dark stars, if they were common, should have similar properties as conventional Pop.~III stars. For MS-dominated models, this requires that typical dark star masses are of order $100\ M_\odot$ or below. For CD models it requires a dark matter density {above} $10^{11}-10^{12}\ \mathrm{GeV}\ \mathrm{cm}^{-3}$ if a spin-dependent elastic scattering cross section of $t\times10^{-39}$~cm$^2$ is assumed \citep{Yoon08}. 
\item {Alternatively, it may imply that the elastic scattering cross section is smaller than the current upper limits, that the dark matter cusp is destroyed by mergers or friction with the gas or that the star is displaced from the center of the cusp.}
\item A further interpretation is that dark stars are very rare. This would require some mechanism to prevent dark star formation in most minihalos.
\item However, if the double-reionization models are actually true, it would indicate that dark stars form only at redshifts beyond $14$, which makes direct observations difficult.
\item We also note that $21$ cm observations may either confirm or rule out double-reionization models.
\end{itemize}
We have also examined whether the formation of dark stars and the corresponding enhancement of dark matter density in dark matter halos due to adiabatic contraction may increase the observed X-ray, gamma-ray and neutrino background. Here we found the following results:
\begin{itemize}
\item {For massive dark matter particles, direct annihilation into gamma-rays provides significant constraints for masses less than $30$~GeV.}
\item {For massive dark matter particles, the contribution from direct annihilation into neutrinos is well below the observed background.}
\item {In light dark matter scenarios, the $511$ keV emission is significantly enhanced below frequencies of $100$ keV in the observers restframe. For a certain range of parameters, this emission may even form a significant contribution of the total X-ray background. In this case, we derive a lower limit of $10$ MeV for the dark matter particle mass (while we find $7$ MeV for standard NFW profiles). }
\item {In light dark matter scenarios, the background radiation due to internal bremsstrahlung is not affected significantly from adiabatic contraction at early times, as the main contribution comes from low redshift. \\}
\item {Both for light and massive dark matter particles, the annihilation products in the remnants of dark stars may provide significant contributions {that may be used to constrain such models in more detail. However, whether this contribution can be reached is highly model-dependent and relevant questions regarding the death of dark stars has not been explored in the literature.}}
\end{itemize}
Future observations may provide further constraints on this exciting suggestion. Small-scale $21$ cm observations may directly probe the HII regions of the first stars and provide a further test of the luminous sources at high redshift, and extremely bright stars might even be observed with the James-Webb telescope, if they form sufficiently late. With this work, we would like to initiate a discussion on observational tests and constraints on dark stars, which may tighten theoretical dark star models and provide a new link between astronomy and particle physics.

\begin{acknowledgments}
We thank Katie Freese for raising our interest in this research during her visit in Heidelberg, and Fabio Iocco for interesting comments and discussions. We also thank Kyungjin Ahn and Simon Glover for interesting discussions on dark matter annihilation and the gamma-ray background, Duane Gruber for providing the HEAO and Comptel-data and Ken Watanabe for providing the SMM data. We acknowledge discussions with Arthur Hebecker on X-ray emission from dark star remnants. DS thanks the Heidelberg Graduate School of Fundamental Physics (HGSFP) and the LGFG for financial support. The HGSFP is funded by the Excellence Initiative of the German Government (grant number GSC 129/1). RB is funded by the
Emmy-Noether grant (DFG) BA 3607/1.  RSK thanks for support from the Emmy Noether grant KL 1358/1. All authors also acknowledge subsidies from the DFG SFB 439 {\em Galaxies in the Early Universe}. {We thank the anonymous referee for {very valuable comments and interesting suggestions }that helped to improve the manuscript.}
\end{acknowledgments}

\clearpage




\clearpage

\end{document}